\documentclass[aps,prb,superscriptaddress,twocolumn,10pt,floatfix]{revtex4-1}
\setcitestyle{numbers,square}
\usepackage[utf8]{inputenc}
\usepackage{amssymb}
\usepackage{amsmath}
\usepackage{multirow}

\usepackage[pdftex]{graphicx}
\usepackage{xspace}
\usepackage{dsfont}
\usepackage{bm}
\usepackage[normalem]{ulem}
\usepackage{hyperref}
\hypersetup{
    colorlinks,%
    citecolor=blue,%
    filecolor=blue,%
    linkcolor=blue,%
    urlcolor=blue
}

\providecommand{\mean}[1]{\langle #1 \rangle} 

\providecommand{\PSHp}{PSH$^+$\xspace}
\providecommand{\PSHm}{PSH$^-$\xspace}

\usepackage{color}

\begin{document}
\title{Spin drift and diffusion in one- and two-subband helical systems}
\author{Gerson J. Ferreira}
\affiliation{Instituto de Física, Universidade Federal de Uberlândia, Uberlândia 38400-902, Minas Gerais, Brazil}

\author{Felix G. G. Hernandez}
\affiliation{Instituto de Física, Universidade de São Paulo, São Paulo 05508-090, São Paulo, Brazil}

\author{Patrick Altmann}
\author{Gian Salis}
\affiliation{IBM Research–Zurich, Säumerstrasse 4, 8803 Rüschlikon, Switzerland}

\date{\today}

\begin{abstract}
The theory of spin drift and diffusion in two-dimensional electron gases is developed in terms of a random walk model incorporating Rashba, linear and cubic Dresselhaus, and intersubband spin-orbit couplings. The additional subband degree of freedom introduces new characteristics to the persistent spin helix (PSH) dynamics. 
As has been described before,
for negligible intersubband scattering rates, the sum of the magnetization of independent subbands leads to a checkerboard pattern of crossed PSHs with long spin lifetime. 
For strong intersubband scattering we model the fast subband dynamics as a new random variable, yielding a dynamics set by averaged spin-orbit couplings of both subbands. In this case the crossed PSH becomes isotropic, rendering circular (Bessel) patterns with short spin lifetime.
Additionally, a finite drift velocity breaks the symmetry between parallel and transverse directions, distorting and dragging the patterns. We find that the maximum spin lifetime shifts away from the PSH regime with increasing drift velocity. We present approximate analytical solutions for these cases and define their domain of validity. Effects of magnetic fields and initial package broadening are also discussed. 
\end{abstract}
\maketitle

\section{Introduction}

Brownian motion \cite{brown1828xxvii,einstein1905molekularkinetischen} provides an elegant description of diffusion processes. A simplified model can be elaborated as a trajectory that consists of successive random steps, where the step size and direction vary according to a given statistical distribution. The extension to spin drift and diffusion \cite{Yang2010RandomWalk} of such a random walk model is a powerful tool to describe the spin dynamics in solid state systems. Spin drift and diffusion can also be described in terms of the quasi-classical kinetic equation \cite{Stanescu2007PhysRevB, Kleinert2007DriftB1, Sinova2012PRB, shen2014theory} and Monte Carlo simulations \cite{ohno2008datta,walser2012direct}.

Tuning the spin-orbit couplings (SOCs) in a two-dimensional electron gas (2DEG) due to structural (Rashba) and bulk (Dresselhaus) inversion asymmetry is an intensely studied method for the coherent control of spin dynamics, which is the motivation \cite{datta1990electronic} and one of the central goals of spintronics research \cite{wolf2001spintronics, RevModPhys.76.323, awschalom2007challenges, behin2010proposal}. Since the initial proposal of the ballistic spin transistor \cite{datta1990electronic}, generalizations have been developed to make it robust against spin-independent scattering  \cite{Schliemann2003spinFET, ohno2008datta, chuang2015all, Souma2015SpinBlockerDoubleWell}, in order to preserve the spin at a certain orientation.

Persistent spin helix (PSH) states were shown to exhibit long spin lifetimes even in the presence of cubic Dresselhaus SOC \cite{bernevig2006exact, koralek2009emergence, Sinova2011StrongSOC, Luffe2011PhysRevB.84.075326, walser2012direct, schliemann2016persistent}. The PSH was first experimentally observed via transient spin-grating spectroscopy  \cite{Weber2007PSHLifetime,koralek2009emergence}. Time-resolved Kerr rotation experiments successfully mapped the diffuse dynamics of optically pumped spin packets  \cite{walser2012direct, Chen2014PhysRevB.90.121304, Altmann2015PhysRevB.92.235304, altmann2016current, kunihashi2016drift} in the PSH regime. Lateral confinement was shown \cite{Altmann2014PhysRevB.90.201306, Altmann2015PhysRevB.92.235304} to further suppress spin decay by restricting the diffusion to one dimension. The Rashba SOC can be controlled via gate voltages \cite{Nitta1997PhysRevLett.78.1335, Studer2009PhysRevLett.103.027201} to achieve or fine tune the PSH regime \cite{Kohda2012PhysRevB.86.081306, ishihara2013direct}. Signatures of the PSH regime and measurements of the SOC were also investigated in weak-localization measurements \cite{sasaki2014direct} and Raman scattering \cite{Raman2014PhysRevB.89.085406}.

The effects of a drift field \cite{Kleinert2007DriftB1, Yang2010RandomWalk} on the dynamics of the PSH states were recently observed \cite{altmann2016current}, allowing a direct measurement of the cubic Dresselhaus coupling $\beta_3$. More recently, for two-subband systems, it was shown \cite{Fu2015Skyrmion} that a crossed-PSH regime (Rashba SOC with opposite sign in each subband) leads to nontrivial spin patterns, which may lead to a topological Hall effect \cite{binz2008chirality}.

In this paper we extend the random walk model for spin drift and diffusion \cite{Yang2010RandomWalk} to incorporate effects of an external magnetic field $\bm{B}$ and two subbands, including the intersubband spin-orbit couplings \cite{bernardes2006spin, EsmerindoPRL2007, calsaverini2008intersubband, dettwiler2017Stretchable, fu2015spin, Fu2015Skyrmion, Souma2015SpinBlockerDoubleWell} ($\Gamma$ and $\eta$), as well as the usual intrasubband Rashba ($\alpha$) and linear ($\beta_1$) and cubic ($\beta_3$) Dresselhaus terms for $[001]$-oriented 2DEG in zinc-blende semiconductors (e.g., GaAs). 
We identify two possible scenarios regarding the intersubband scattering (ISS) rate. For weak ISS, the subbands are effectively uncoupled, yielding independent ensembles. The precession pattern is given by an incoherent sum of the magnetization of the individual subbands, which shows a checkerboard pattern with long spin lifetime in the crossed-PSH regime, in agreement with Ref.~\onlinecite{Fu2015Skyrmion}.
On the other hand, for strong ISS, the resulting dynamics is dominated by subband-averaged SOCs, which will be driven out of the PSH regimes by the fast subband dynamics, yielding a circular (Bessel) pattern with short spin lifetime. 

Before discussing the two-subband systems, we first revisit the single-subband random walk model \cite{Yang2010RandomWalk} to investigate the effects of magnetic fields, drift velocity, and the initial broadening of optically pumped spin packets. 
We show that a finite in-plane drift field (e.g., along $y \parallel [110]$) leads to distinct precession patterns and relaxation rates for the PSH regimes $\alpha = \pm(\beta_1-\beta_3)$, which we label as \PSHp and \PSHm, respectively. 
Our \PSHp solution matches previous discussions in the literature \cite{bernevig2006exact, koralek2009emergence, Stanescu2007PhysRevB, Yang2010RandomWalk, altmann2016current}.
More interestingly, for the \PSHm regime, the drift velocity shifts the maximum spin lifetime away from the precise PSH tuning.
The resulting precession pattern is also strongly affected by the initial broadening of the spin packet. We derive analytical solutions for the narrow and wide packet limits and compare with numerical simulations.
Additionally, we show that a magnetic field combined with a finite drift velocity adds corrections to both the precession frequency and the spin pattern wavelength.

This paper is organized as follows. In Sec.~\ref{sec:rwmodel} we introduce the random walk model to establish the notation and identify its main aspects and limitations. Next, in Sec.~\ref{sec:Single}, we discuss in detail the diffusive dynamics for the single-subband case. We present analytical approximate solutions valid for a wide range of parameters near the PSH regimes. These are compared with exact numerical solutions. Here we also discuss the expected effects of finite magnetic fields. The two-subband case is discussed in Sec.~\ref{sec:twosubbands}. We consider a two-subband 2DEG with intraband Rashba and Dresselhaus SOCs, as well as intersubband SOC. We close the paper with general remarks and the conclusions.

\section{Random Walk For Spin Diffusion}
\label{sec:rwmodel}

The random walk \cite{brown1828xxvii,einstein1905molekularkinetischen,Yang2010RandomWalk} (RW) is characterized by the random motion of a particle, which here is an electron that scatters via different processes (e.g., impurity sites, defects, electron-electron scattering, phonons). In between scattering events the electron ballistically travels a distance $\Delta\bm{r} = \bm{v}\tau_r$, where both the velocity $\bm{v}$ and the scattering time $\tau_r$ are random variables. Here, $\bm{v} = v_F \hat{\theta} + \bm{v}_d$, where $\hat{\theta}$ is a uniformly random direction (along the $xy$ plane), $v_F$ is the Fermi velocity, and $\bm{v}_d = \tau e \bm{E}/m$ is the drift velocity due to the electric field $\bm{E}$. Since the scattering events are independent, the scattering time $\tau_r$ is expected to obey a Poissonian distribution, such that $\mean{\tau_r} = \tau$ and $\mean{\tau_r^2} = 2\tau^2$.

Throughout the ballistic motion, the electron spin precesses due to external magnetic fields or internal velocity-dependent spin-orbit fields. The average scattering time $\tau$ is considered to be short compared with the spin precession period, which allows us to describe below the ballistic spin evolution perturbatively. Considering only non-magnetic scattering, the spin is preserved at each collision, but its precession direction changes due to the SOC.
This description leads to a model that is consistent with the Dyakonov-Perel dynamics, which is adequate for typical semiconductors, e.g., GaAs, where this is the dominant mechanism for spin decay.

Let us start with a discrete time dynamics labeled by the step index $n$. The position of the electron at the time step $n+1$ is then $\bm{r}_{n+1} = \bm{r}_n + \bm{v}_n \tau_n$. The velocity $\bm{v}_n$ and the time interval $\tau_n$ depend on the step $n$ as they are randomized at each collision. During $\tau_n$ the motion is ballistic and the spin evolution is given as

\begin{equation}
 \dfrac{\partial}{\partial t} \bm{s} = \bm{\Omega} \times \bm{s}.
\end{equation}
Typically, the precession term $\bm{\Omega}$ is given by external magnetic fields and the SOCs. But for now let us keep it arbitrary, with the only constraint that the equation above is linear in $\bm{s}$. An approximate solution for $|\Omega_n\tau_n| \ll 1$ is obtained iterating the equation above up to second order, yielding

\begin{equation}
 \bm{s}_{n+1} \approx \bm{s}_n + \tau_n \bm{\Omega}_n\times\bm{s}_n + \dfrac{\tau_n^2}{2}\bm{\Omega}_n \times (\bm{\Omega}_n \times \bm{s}_n).
 \label{eq:ssol}
\end{equation}
Here $\bm{\Omega}_n \equiv \bm{\Omega}(\bm{v}_n)$ varies in each time step because the SOC depends on $\bm{v}_n$.

For an ensemble of spins, the magnetization profile $\bm{m}_{n+1}(\bm{r})$ at time step $n+1$ can be written in terms of a joint probability $P_{n+1}(\bm{r}; \bm{s})$ of finding an electron at time step $n+1$ at position $\bm{r}$ having spin $\bm{s}$,

\begin{equation}
 \bm{m}_{n+1}(\bm{r}) = \int \bm{s} P_{n+1}(\bm{r}; \bm{s}) d\Sigma,
 \label{eq:mint}
\end{equation}
where the integral runs over the Bloch sphere. Since the scattering process is random, the joint probability can be written as the average result of all possible paths from $n$ to $n+1$ that lead to an electron at $\bm{r}$ with spin $\bm{s}$,

\begin{equation}
 P_{n+1}(\bm{r}; \bm{s}) = \mean{ P_{n}(\bm{r}-\bm{v}_{n}\tau_n; \bm{s}-\Delta \bm{s}_{n}) },
 \label{eq:joint}
\end{equation}
where $\Delta \bm{s}_{n} = \bm{s}_{n+1}-\bm{s}_n$, and 
$\mean{\cdots}$ denotes the average over the momentum direction $\hat{\theta}$ and the scattering time $\tau_n$.

To recover a differential equation for $\bm{m}(\bm{r},t)$, one expands the average above around $\mean{ P_{n}(\bm{r}; \bm{s}) }$ up to second order in $\bm{v}_n \tau_n$ and zero order in $\Delta\bm{s}_n$. Combining all expressions above and converting the discrete time back to the continuum, we get

\begin{align}
\nonumber
 \dfrac{\partial}{\partial t} \bm{m}(\bm{r},t) &= \Big(\Lambda_{dd} + \Lambda_{pr}\Big)\bm{m}(\bm{r},t),\\
\label{eq:diffmag}
 \Lambda_{dd} &= -\bm{v}_d\cdot\bm{\nabla} + \tau\nabla^2_v,\\
\nonumber
 \Lambda_{pr} &= 
 \begin{pmatrix}
  - \tau\mean{\Omega_y^2} & \tau\mean{\Omega_x \Omega_y} & \Xi_y\\
  \tau\mean{\Omega_x \Omega_y} & - \tau\mean{\Omega_x^2} & -\Xi_x\\
  -\Xi_y & \Xi_x & -\tau\mean{\Omega_x^2}-\tau\mean{\Omega_y^2}
 \end{pmatrix}
\end{align}
where the diagonal term $\Lambda_{dd}$ drives the drift and diffusion process, while the matrix $\Lambda_{pr}$ dictates the spin precession and relaxation. The new terms above read

\begin{align}
 \nabla^2_v &= \mean{v_x^2}\partial^2_x + \mean{v_y^2}\partial^2_y,\\
 \Xi_x &= \mean{\Omega_x} - 2\tau\Big[ \mean{v_x\Omega_x}\partial_x + \mean{v_y\Omega_x}\partial_y \Big],\\
 \Xi_y &= \mean{\Omega_y} - 2\tau\Big[ \mean{v_x\Omega_y}\partial_x + \mean{v_y\Omega_y}\partial_y \Big],
\end{align}
where we have assumed $\Omega_z = 0$ for simplicity. This is the case in a $[001]$-oriented 2DEG, where Rashba and Dresselhaus SOC contributions are in-plane. The extra terms for a finite $\Omega_z$ are shown in Appendix \ref{app:omegaz}.

The resulting Eq.~\eqref{eq:diffmag} differs from those of Ref.~\onlinecite{Yang2010RandomWalk} as we consider here the Poissonian distribution of the scattering time, such that $\mean{\tau_n} = \tau$ and $\mean{\tau_n^2} = 2\tau$. Moreover, we keep Eq.~\eqref{eq:diffmag} in a general form that will allow us to include the external magnetic field and consider two subbands.

\subsection{Numerical implementation and q-space}

The averages that define Eq.~\eqref{eq:diffmag} are simple expressions of the system parameters (see next section and the appendices). Therefore, the only numerical task remaining is to properly solve the initial value problem. Applying a spatial Fourier transform ($\bm{r} \rightarrow \bm{q}$), the derivatives become $\partial_{x/y} \rightarrow -i q_{x/y}$, and the solution in $q$ space is simply

\begin{equation}
 \bm{\tilde{m}}(\bm{q},t) = e^{\tilde{\Lambda} t}\bm{\tilde{m}}(\bm{q},0),
 \label{eq:mkspace}
\end{equation}
where $\bm{\tilde{m}}(\bm{q},0)$ is the Fourier transform of the initial packet, and $\tilde{\Lambda}$ is the Fourier transform of the matrices $\Lambda_{dd}$ and $\Lambda_{pr}$ in Eq.~\eqref{eq:diffmag}. Namely, $\tilde{\Lambda}$ is obtained with the replacements: $-\bm{v}_d\cdot\bm{\nabla} \rightarrow i \bm{v}_d\cdot\bm{q}$, $\nabla^2_v \rightarrow \tilde{\nabla}^2_v = -(\mean{v_x^2}q_x^2 + \mean{v_y^2}q_y^2)$, and $\bm{\Xi} \rightarrow \bm{\tilde{\Xi}} = \mean{\bm{\Omega}} + 2i\tau\mean{(\bm{v}\cdot\bm{q})\bm{\Omega}}$. Hereafter we use the symbol $\sim$ to refer to quantities in $q$ space.

The matrix exponential $e^{\tilde{\Lambda} t}$ can be easily calculated in terms of its eigenvalues and eigenvectors. Therefore, the only relevant numerical \cite{*[{The numerical calculations are developed using the \href{www.julialang.org}{Julia} language [}] [{].}] julia} task is to perform the two-dimensional inverse Fourier transform ($\bm{q} \rightarrow \bm{r}$) at different times $t$ to obtain $\bm{m}(\bm{r},t)$. Since no extra approximations are involved, we shall consider the numerical evolution as exact solutions of Eq.~\eqref{eq:diffmag}.

\subsection{Initial broadening}

In optical pump-probe experiments, the initial magnetization packet is set by the laser spot, which here is characterized by the initial broadening $\Gamma_0$. Therefore, in general we shall consider isotropic Gaussian packets polarized along $z$, i.e., $\bm{m}(\bm{r},0) \propto e^{-\frac{1}{2}(r/\Gamma_0)^2} \hat{z}$, as the initial condition for Eq.~\eqref{eq:diffmag}. For analytical solutions, we also use initial delta-packets, $\bm{m}(\bm{r},0) = \delta(\bm{r})\hat{z}$, which corresponds to the limit $\Gamma_0 \rightarrow 0$.

Intuitively, one would expect that the $\delta$-packet solution could be used to obtain the dynamics of any other initial packet via convolution. However, for the \PSHm regime, we derive two different analytical but approximate solutions of Eq.~\eqref{eq:diffmag} by neglecting distinct terms in $\Lambda_{pr}$. While one approximation is compatible with a narrow packet, the other is appropriate for broad packets. Consequently, since we do not have a general exact solution for a $\delta$ packet, one cannot convolve the approximate $\delta$-packet solution to broad packets.

\section{Single Subband}
\label{sec:Single}

The theoretical analysis of the drift field on the diffuse spin dynamics were first presented in Refs.~\onlinecite{Kleinert2007DriftB1, Yang2010RandomWalk} for a single-subband system, and recently observed experimentally \cite{altmann2016current}. In this section we explore and extend these results using the RW model. We show new analytical solutions for the \PSHm regime ($\alpha \approx -\beta_1 + \beta_3$), and include the effects of external magnetic fields. Additionally, we compare the solutions for spatially wide and narrow initial packets. Away from these limits, we solve the RW model numerically for comparison. 

The Hamiltonian for the single-subband 2DEG is $H = \varepsilon_1 + \frac{\hbar^2}{2m}k^2 + \frac{\hbar}{2} \bm{\sigma}\cdot \bm{\Omega}$, where the SOC is given by the Rashba ($\alpha$) and linear and cubic Dresselhaus ($\beta_1$ and $\beta_3$) terms as

\begin{equation}
  \bm{\Omega} = \dfrac{2}{\hbar}
 \begin{pmatrix}
  (+\alpha+\beta_{1})k_y + 2\beta_{3}\dfrac{k_x^2-k_y^2}{k^2}k_y\\
  (-\alpha+\beta_{1})k_x - 2\beta_{3}\dfrac{k_x^2-k_y^2}{k^2}k_x\\
  0
 \end{pmatrix}.
 \label{eq:OmegaSOC}
\end{equation}
Here $x \parallel [1\bar{1}0]$ and $y\parallel [110]$.
Treating the SOC as a weak perturbation to the band structure, the velocity is simply $\bm{v}(\bm{k}) = \hbar \bm{k}/m$. Therefore the averages $\mean{\cdots}$ over the random motion direction $\hat{\theta}$ shall be read as an average of $\bm{k}$ over the Fermi circle $|\bm{k}| = k_F$. Considering the drift velocity $\bm{v}_d = v_d \hat{y}$ we find $\mean{\Omega_y} = \mean{\Omega_x \Omega_y} = \mean{v_x\Omega_x} = \mean{v_y\Omega_y} =  0$. The other averages remain finite and are shown in Appendix \ref{app:single}. Within this section we will use the parameters shown in Table \ref{tab:single}.

\begin{table}[hb!]
\caption{Parameters considered for the single-subband discussion. The value of the Rashba coefficient $\alpha$ varies from the \PSHm to the \PSHp regime in the range $-(\beta_1-\beta_3) \leq \alpha \leq (\beta_1-\beta_3)$. The cubic Dresselhaus term near the Fermi level is $\beta_3 \approx \gamma\pi n_s/2$, and $\gamma = 11$~eV\AA$^3$ is the bulk Dresselhaus coefficient.
}
\label{tab:single}
\begin{center}
\begin{ruledtabular}
\begin{tabular}{cll}
Parameter & Value & Description\\
\hline
$m$       & $0.067m_0$ & Effective mass (GaAs)\\
\multicolumn{2}{c}{$-(\beta_1-\beta_3) \leq \alpha \leq (\beta_1-\beta_3)$} & Rashba SOC\\
$\beta_1$ & 3.7~meV\AA & Linear Dresselhaus SOC\\
$\beta_3$ & 0.7~meV\AA & Cubic Dresselhaus SOC\\
$n_s$     & $4\times 10^{11}$~cm$^{-2}$ & 2DEG density\\
$\tau$    & 1~ps & Average scattering time\\
\end{tabular}
\end{ruledtabular}
\end{center}
\end{table}

To go forward and find analytical solutions of Eq.~\eqref{eq:diffmag} we must make approximations. We will consider the \PSHp ($\alpha \approx \beta_1-\beta_3$) and \PSHm ($\alpha \approx -\beta_1+\beta_3$) regimes. These regimes are different because we keep the drift velocity fixed along $\hat{y}$. Equivalently, for a fixed set of SOC coefficients one could alternate between the PSH$^\pm$ regimes switching the drift velocity direction \cite{altmann2016current} between $\hat{x}$ and $\hat{y}$.  

\begin{figure}[ht!]
 \centering
 \includegraphics[width=\columnwidth,keepaspectratio=true]{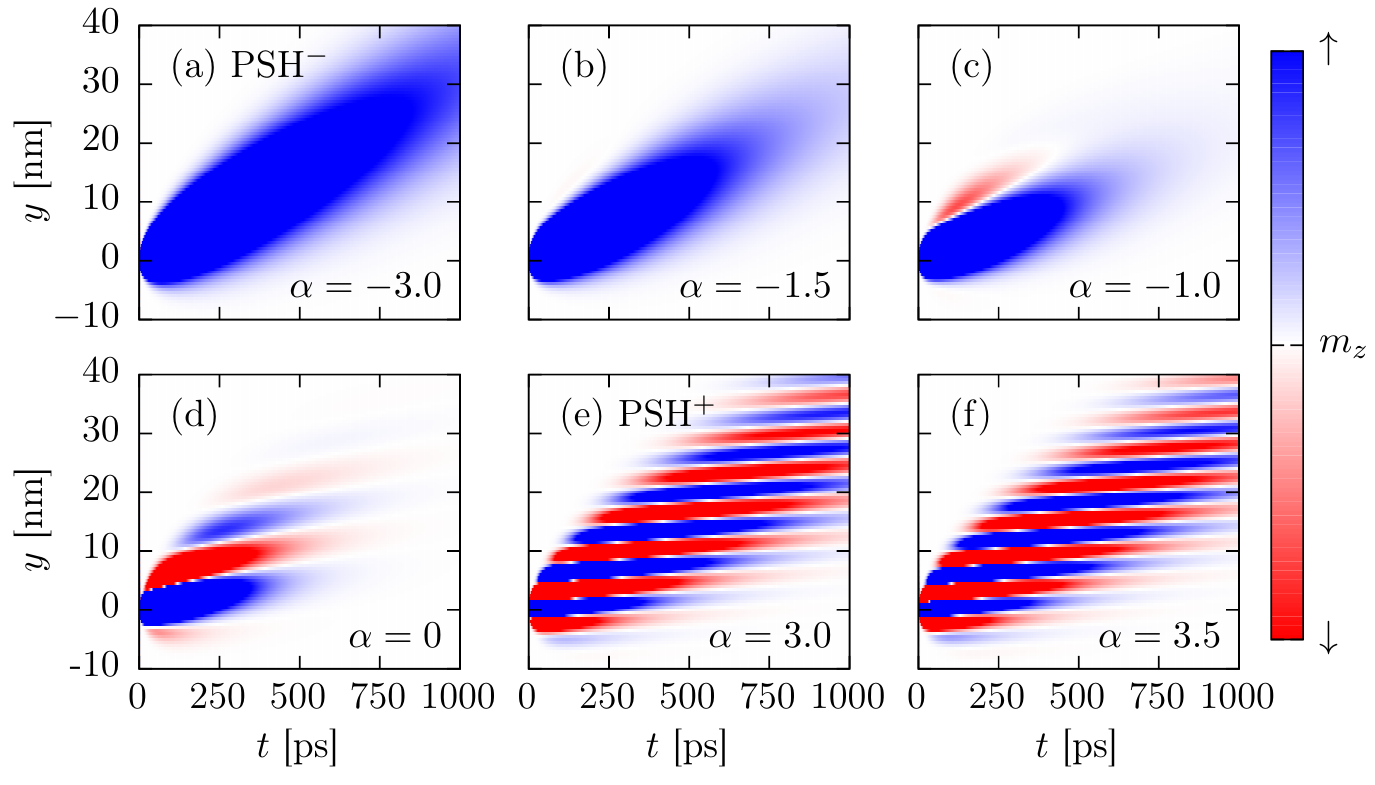}
 \caption{Transition from \PSHm to \PSHp as a function of $\alpha$ for an initially narrow packet, $m_z(\bm{r},0) = \delta(\bm{r})$. Each panel is for a different value of $\alpha$, which ranges from the (a) \PSHm to (e) \PSHp regime. The labels indicate the value of $\alpha$ in meV\AA.
 These and the following images are saturated for better visualization of the magnetization patterns.}
 \label{fig:narrowvsalpha}
\end{figure}

\subsection{\texorpdfstring{\PSHp: $\alpha \approx \beta_1-\beta_3$}{PSH+}}
\label{sec:pshp}

To establish the approximations for the \PSHp regime, let us compare the nondiagonal terms $\tilde{\Xi}_x$ and $\tilde{\Xi}_y$ in the Fourier space of Eq.~\eqref{eq:diffmag}. The intensity of $\tilde{\Xi}_x = \mean{\Omega_x} + 2i\tau q_y \mean{v_y\Omega_x}$ scales with $(\alpha+\beta_1)$, while $\tilde{\Xi}_y = 2i\tau q_x \mean{v_x\Omega_y}$ scales with $(\alpha-\beta_1)$; see Appendix \ref{app:single}. Since our initial packages are always isotropic, the ranges of $q_x$ and $q_y$ are similar, which allow us to approximate both $|q_x|$ and $|q_y| \lesssim 1/\Gamma_0$ to compare the intensities of $\tilde{\Xi}_x$ and $\tilde{\Xi}_y$. For $\alpha \approx \beta_1-\beta_3$ and $\beta_3 \ll \alpha + \beta_1$, we have $|\tilde{\Xi}_x| \gg |\tilde{\Xi}_y|$.

\begin{figure}[ht!]
 \centering
 \includegraphics[width=\columnwidth,keepaspectratio=true]{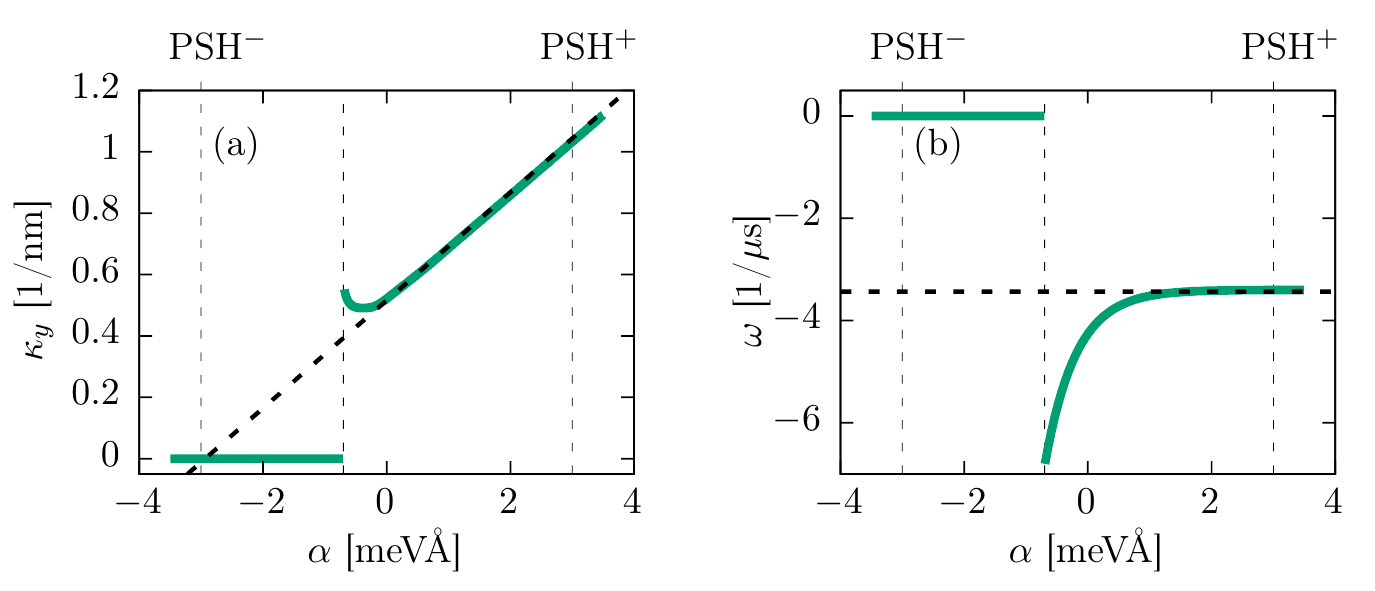}
 \caption{(a) Wave number $\kappa_y$ and (b) frequency $\omega$ extracted from Fig.~\ref{fig:narrowvsalpha}. For $\alpha < -0.7$~meV\AA~ the stripes vanish accompanied by discontinuities in $\kappa_y$ and $\omega$. The \PSHp solutions given by Eq.~\eqref{eq:qyplus} and Eq.~\eqref{eq:wplus}, shown as thick dashed lines, match well the numerical data for $\alpha > 0$.}
 \label{fig:qw}
\end{figure}

We can split the matrix in Eq.~\eqref{eq:diffmag} in two blocks: a one-dimensional block composed of the $m_x(\bm{r},t)$ component only, and a two-dimensional block composed of the remaining components, $m_y(\bm{r},t)$ and $m_z(\bm{r},t)$. These blocks are coupled by $\tilde{\Xi}_y$. If the difference between eigenvalues of these blocks is large compared to their coupling, one can neglect $\tilde{\Xi}_y$. The approximate eigenvalues of the $yz$ subspace are then

\begin{equation}
 \tilde{\lambda}^\pm_{yz} \approx \tilde{\lambda}_0 -\tau\left(\mean{\Omega_x^2}+\dfrac{\mean{\Omega_y^2}}{2}\right) \pm i \, \tilde{\Xi}_x,
\end{equation}
where we have used $|\tau\mean{\Omega_y^2}| \ll |\tilde{\Xi}_x|$, which follows from the scaling of these quantities with $(\alpha \pm \beta_1)$. The eigenvalue of the $x$ subspace is $\tilde{\lambda}_x = \tilde{\lambda}_0-\tau\mean{\Omega_y^2}$. The common diagonal term $\tilde{\lambda}_0 = i v_d q_y -\tau\tilde{\nabla}^2_v$ dictates the drift and diffusion. In terms of these eigenvalues, the condition to neglect the coupling $\tilde{\Xi}_y$ reads $|\tilde{\lambda}_{yz}^\pm - \tilde{\lambda}_x| \gg |\tilde{\Xi}_y|$. This is satisfied near the \PSHp regime, but fails near the \PSHm regime. Therefore we can always neglect $\tilde{\Xi}_y$ near the \PSHp regime, and the precession is dominated by the lower block of the matrix in Eq.~\eqref{eq:diffmag}, corresponding to the $(m_y, m_z)$ subspace. The numerical solutions in Fig.~\ref{fig:narrowvsalpha} and Fig.~\ref{fig:qw} show a transition between these two regimes near $\alpha = -0.7$~meV\AA.

The approximation $\tilde{\Xi}_y \rightarrow 0$ allows us to write the $q$-space solution, Eq.~\eqref{eq:mkspace}, in simple terms and calculate the inverse Fourier transform to obtain the $z$ component of the magnetization profile, which reads

\begin{align}
 \label{eq:PSHplus}
 m_z^+(\bm{r},t) &= \rho(\bm{r},t) e^{-\gamma_p t} \cos(\kappa_y y + \omega t),\\
 \rho(\bm{r},t) &= \dfrac{1}{\Gamma_{x,t}\Gamma_{y,t}} e^{-\frac{x^2}{2\Gamma_{x,t}^2}} \; e^{-\frac{(y-v_d t)^2}{2\Gamma_{y,t}^2}},
\end{align}
where the broadenings are $\Gamma_{x,t}^2 = 2Dt$, $\Gamma_{y,t}^2 = 2(D+\tau v_d^2)t$ and the diffusion coefficient is $D = \tau v_F^2/2$. The term $\rho(\bm{r},t)$ drives the drift and diffusion, and it is common to all following solutions discussed hereafter. The wave vector $\kappa_y$, frequency $\omega$, and relaxation rate $\gamma_p$ are defined by $\Omega$ averages (see Appendix \ref{app:pshp}). 
In terms of the SOC coefficients, up to leading order in $v_d/v_F$, these are

\begin{align}
\label{eq:qyplus}
 \kappa_y &\approx \dfrac{2m}{\hbar^2}\left[ \alpha + \beta_{1} - \beta_{3} -8\beta_3\dfrac{v_d^2}{v_F^2}\right],\\
\label{eq:wplus}
  \omega &\approx -\dfrac{2m}{\hbar^2}v_d\beta_{3}\left(1 - 6\dfrac{v_d^2}{v_F^2}\right),
\end{align}
\begin{multline}
 \gamma_p \approx \dfrac{\tau m^2}{\hbar^4}v_F^2
   \Bigg[
      3 \beta_3^2 + (\alpha-\beta_1+\beta_3)^2 \\
      - 4 \dfrac{v_d^2}{v_F^2} (\alpha - \beta_1 - 9\beta_3)\beta_3
   \Bigg].
   \label{eq:gammap}
\end{multline}

The $\kappa_y$, $\omega$, $\gamma_p$ and $D$ above match those of Refs.~\onlinecite{Salis2014Dynamics, altmann2016current} for small $v_d$. In contrast, our relaxation rate $\gamma_p$ and diffusion constant $D$ are twice those of Ref.~\onlinecite{Yang2010RandomWalk} due to the Poissonian distribution of the scattering time $\tau_r$ considered here.

The resulting pattern of $m_z^+(\bm{r},t)$ is shown in Fig.~\ref{fig:narrowvsalpha}(e), where we compare it with the numerical solutions beyond the PSH$^\pm$ regimes. The stripes of oscillating spins constitute a magnetization wave moving along $y$ with velocity $-\omega/\kappa_y \approx v_d\beta_3/2\beta_1$, and an envelope profile $\rho(\bm{r},t)$. Figure \ref{fig:qw} shows that the analytical solution above is valid over a wide range of $\alpha$ beyond the \PSHp regime.

\begin{figure}[ht!]
 \centering
 \includegraphics[width=\columnwidth,keepaspectratio=true]{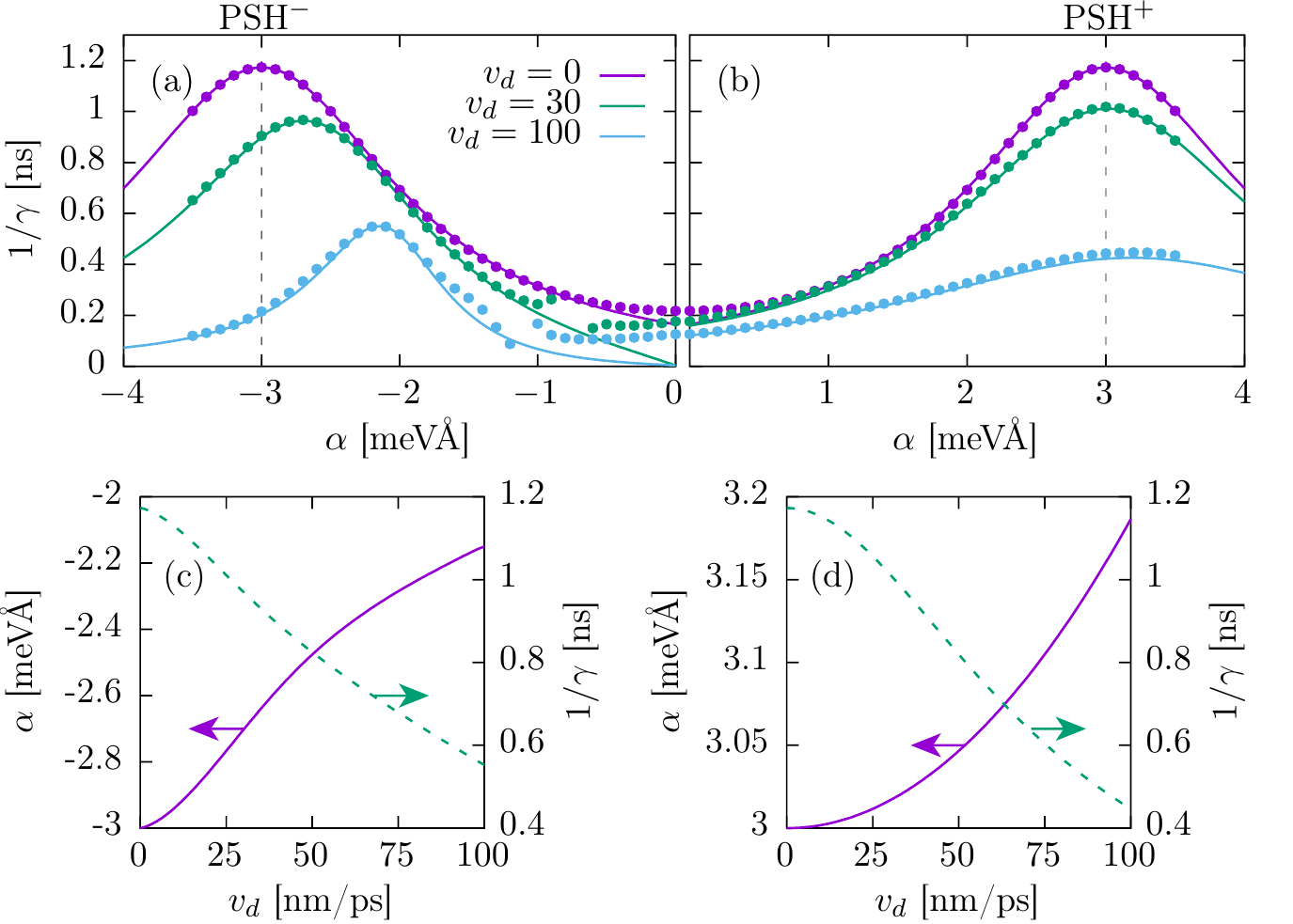}
 \caption{(a), (b) Decay time $1/\gamma$ as a function of $\alpha$ for different $v_d$ (in nm/ps). 
 The circles were extracted as exponential fits to the numerical solutions in Fig.~\ref{fig:narrowvsalpha}.
 The analytical solutions (solid lines) match the numerical data for $\alpha<0$ using the \PSHm $1/\gamma_n$ [Eq.~\eqref{eq:gamman}], while for $\alpha > 0$ it matches the \PSHp $1/\gamma_p$ [Eq.~\eqref{eq:gammap}].
  (c), (d) Peak position ($\alpha$) and intensity ($1/\gamma$) as a function of $v_d$ for the PSH$^\pm$ peaks, respectively.
 }
 \label{fig:gamma}
\end{figure}

Figures \ref{fig:gamma}(a) and \ref{fig:gamma}(b) show the decay time $1/\gamma_p$ as a function of $\alpha$, comparing the analytical solution of Eq.~\eqref{eq:gammap} with the numerical simulations for different $v_d$. The precise agreement validates the approximations above. With increasing $v_d$, the peak of maximum lifetime shifts to larger $\alpha$ and looses intensity, as seen in Fig.~\ref{fig:gamma}(d). This effect is more pronounced for the \PSHm regime, which we discuss next.

\subsection{\texorpdfstring{\PSHm: $\alpha \approx -\beta_1+\beta_3$}{PSH-}}
\label{sec:pshm}

While the \PSHp regime was already introduced in Ref.~\onlinecite{Yang2010RandomWalk}, in this section we show that the \PSHm regime presents novel solutions for the random walk problem. Particularly, the spin precession pattern in this regime is sensitive to the initial package broadening, and the relaxation rate strongly depends on the drift velocity.

\begin{figure}[ht!]
 \centering
 \includegraphics[width=\columnwidth,keepaspectratio=true]{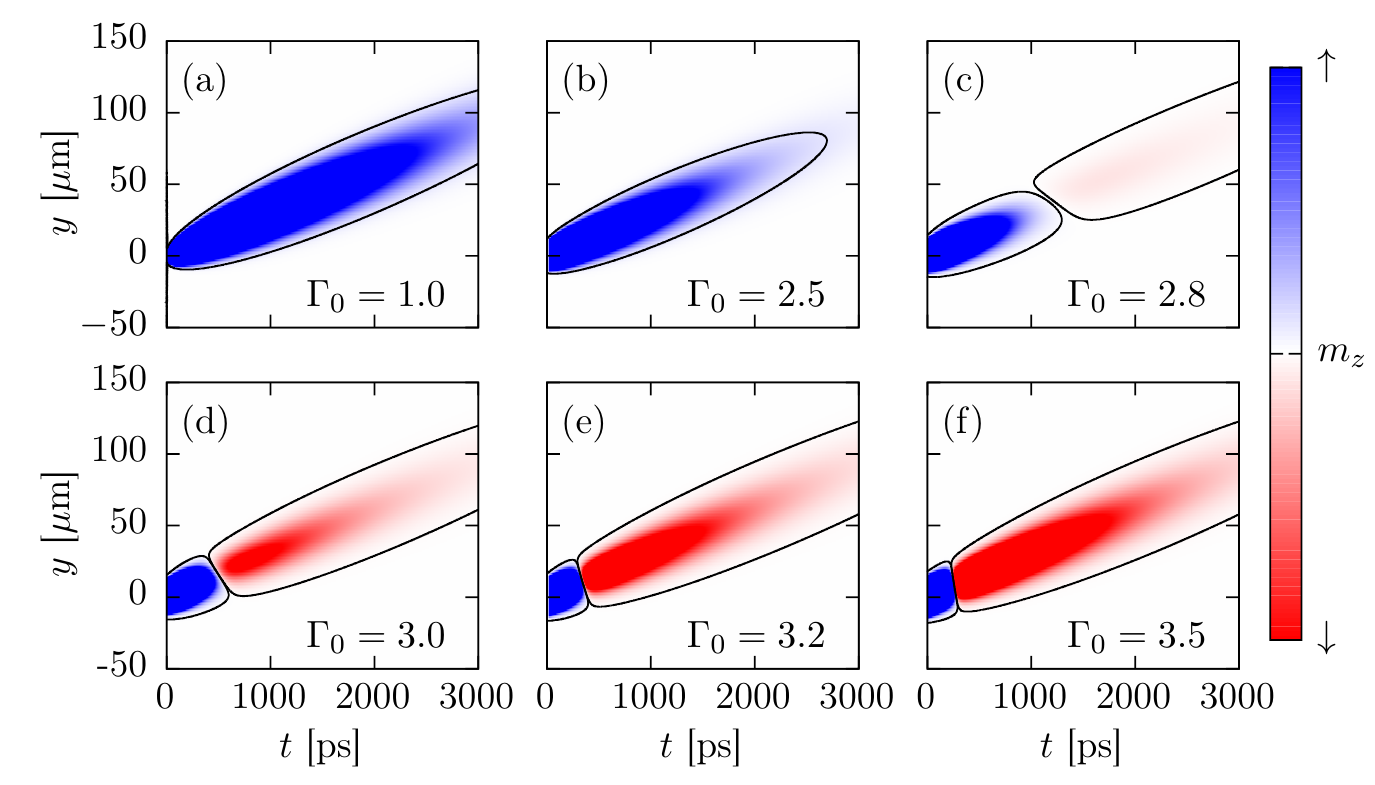}
 \caption{Magnetization pattern at the \PSHm regime for initially wide packets for different initial broadenings $\Gamma_0$, as indicated in each panel in $\mu$m. Figure \ref{fig:narrowvsalpha}(a) corresponds to the $\Gamma_0 \rightarrow 0$ limit. The spin flip only occurs for $\Gamma_0 \gtrsim \Gamma_C \approx 2$~$\mu$m. In panel (a), for $\Gamma_0 = 1$~$\mu$m, the magnetization fades away for $t > 3000$~ps, but does not flip, while in panel (b) a node forms around $t = 3000$~ps. From panels (c) to (f) the transition instant $t_c$ shifts towards smaller $t$ with increasing $\Gamma_0$. Equation \eqref{eq:nodetime} defines $t_c \approx 100$~ps in the limit $\Gamma_0 \rightarrow \infty$. Black line contours are guides to the eye.}
 \label{fig:PSHminus-vsbroad}
\end{figure}

For the \PSHm regime, $\alpha \approx -\beta_1+\beta_3$, we get now $|\tilde{\Xi}_y| \gg |\tilde{\Xi}_x|$, due to their scaling with $(\alpha \pm \beta_1)$. This suggests a splitting of the matrix in Eq.~\eqref{eq:diffmag} into an $xz$ block weakly coupled to the $x$ term by $\tilde{\Xi}_x$. However, one can only consider the coupling to be weak if the difference between the eigenvalues of the blocks is much bigger than their coupling. Noticing that $\mean{\Omega_x^2} \ll \mean{\Omega_y^2}$, the eigenvalues of the uncoupled blocks ($\tilde{\Xi}_x = 0$) are

\begin{align}
 \tilde{\lambda}_y &= \tilde{\lambda}_0 -\tau \mean{\Omega_x^2},\\
 \tilde{\lambda}_{xz}^\pm &\approx \tilde{\lambda}_0 -\tau\mean{\Omega_y^2} \pm i\,\tilde{\Xi}_y.
\end{align}
Therefore, the decoupling condition becomes $|\tilde{\lambda}_y - \tilde{\lambda}_{xz}^\pm| \gg |\tilde{\Xi}_x|$. As in the \PSHp regime, the range of $q_x$ and $q_y$ is about $|q_x| = |q_y| \lesssim 1/\Gamma_0$, which we use to estimate $\tilde{\Xi}_x$ and $\tilde{\Xi}_y$. We find two possible scenarios to satisfy the decoupling: (i) narrow initial packages, $\Gamma_0 \ll \Gamma_C$, and (ii) wide initial packages, $\Gamma_0 \gg \Gamma_C$. The critical initial broadening $\Gamma_C$, around which the transition occurs, is

\begin{equation}
 \Gamma_C \approx \dfrac{2 \mean{v_x\Omega_y}}{\mean{\Omega_y^2}} \approx \dfrac{\hbar^2}{2m\beta_1} \left[ 1 + \dfrac{\beta_3}{\beta_1} \right].
\end{equation}
For the set of parameters used here we find $\Gamma_C \approx 1.8$~$\mu$m; see Fig.~\ref{fig:PSHminus-vsbroad}. Next, we discuss the narrow ($\Gamma_0 \ll \Gamma_C$) and wide ($\Gamma_0 \gg \Gamma_C$) initial packet cases separately.

\subsubsection{\texorpdfstring{Initially narrow packet: $\Gamma_0 \ll \Gamma_C$}{Initially narrow packet}}
\label{sec:narrow}

For $\Gamma_0 \ll \Gamma_C$ the precession is set by the $xz$ block of Eq.~\eqref{eq:diffmag}, since $|\tilde{\Xi}_y| \propto 1/\Gamma_0$ in $\tilde{\lambda}_{xz}^\pm$ becomes large. Within this subspace, we can solve Eq.~\eqref{eq:diffmag} analytically to find

\begin{equation}
 m_z^-(\bm{r},t) =
 \rho(\bm{r},t)
 e^{-\gamma_n t} \cos(\kappa_x x),
 \label{eq:PSHminusNarrow}
\end{equation}
where $\rho(\bm{r},t)$ is the same drift-diffusion term from the \PSHp regime, $\gamma_n$ is the relaxation rate, and the wave number $\kappa_x = \mean{v_x\Omega_y}/\mean{v_x^2}$, which up to leading order in $v_d/v_F$ reads

\begin{equation}
 \kappa_x \approx \dfrac{2m}{\hbar^2}\left[-\alpha+\beta_1-\beta_3+2\beta_3 \dfrac{v_d^2}{v_F^2}\right].
 \label{eq:qxminus}
\end{equation}
This magnetization profile is shown in Fig.~\ref{fig:narrowvsalpha}(a) as a function of $y$ and time $t$ for $x=0$ and $v_d = 30$~nm/ps. In accordance to the equation above, there is no precession along $y$.

For typical parameters we find that, although small, the coupling $\tilde{\Xi}_x \approx \mean{\Omega_x} \propto v_d$ has to be included to properly describe $\gamma_n$ for finite $v_d$. We include this coupling into the solution using second-order perturbation theory to correct the eigenvalue $\tilde{\lambda}_{xz}^\pm \rightarrow \tilde{\lambda}_{xz}^\pm + \delta\tilde{\lambda}_{xz}$, where

\begin{equation}
 \delta\tilde{\lambda}_{xz} \approx \dfrac{1}{2\tau}\dfrac{\mean{\Omega_x}^2}{\mean{\Omega_y^2}-\mean{\Omega_x^2}},
\end{equation}
and the resulting relaxation rate reads

\begin{equation}
 \gamma_n \approx \tau\left(\mean{\Omega_y^2}+\dfrac{\mean{\Omega_x^2}}{2}-\dfrac{\mean{v_x\Omega_y}^2}{\mean{v_x^2}}\right) + \delta\tilde{\lambda}_{xz},
 \label{eq:gamman}
\end{equation}
which strongly depends on the drift velocity as shown in Figs.~\ref{fig:gamma}(a)-\ref{fig:gamma}(c). The resulting expression for $\gamma_n$ in terms of the SOCs is long (not shown). In contrast to the \PSHp regime, for increasing $v_d$ the \PSHm peak of maximum lifetime strongly shifts away from $\alpha = -(\beta_1-\beta_3)$.

\subsubsection{\texorpdfstring{Initially wide packet: $\Gamma_0 \gg \Gamma_C$}{Initially wide packet}}
\label{sec:wide}

For $\Gamma_0 \gtrsim \Gamma_C$ we cannot split the matrix in Eq.~\eqref{eq:diffmag} into simpler blocks. Here, both $\tilde{\Xi}_x$ and $\tilde{\Xi}_y$ are relevant. However, a qualitative description can be found in the limit $\Gamma_0 \rightarrow \infty$, such that $\tilde{\Xi}_x \rightarrow \mean{\Omega_x}$ and $\tilde{\Xi}_y \rightarrow 0$. The precession dynamics is given by the $yz$ block. Although small, the only relevant coupling remaining is $\tilde{\Xi}_x \approx \mean{\Omega_x} \propto v_d$. Consequently, a small, yet finite drift velocity is required to observe this regime.

Within these approximations, it is easy to solve Eq.~\eqref{eq:diffmag} in Fourier space and return with the inverse transform to obtain

\begin{multline}
\label{eq:PSHminusBroad}
 m_z^-(\bm{r}, t) = \rho(\bm{r},t) e^{-\gamma_w t}\\
 \times \left[ \cosh(\xi t) - \dfrac{\tau\mean{\Omega_y^2}}{2\xi}\sinh(\xi t) \right].
\end{multline}
Here the broadenings in $\rho(\bm{r},t)$ are approximately constant, $\Gamma_{x,t} = \Gamma_{y,t} \approx \Gamma_0$, due to the large initial broadening $\Gamma_0$. The relaxation rate $\gamma_w$ and hyperbolic frequency $\xi$ are shown in terms of the $\Omega$ averages in Appendix \ref{app:pshm}. For $\alpha \approx -(\beta_1-\beta_3)$ and up to leading order in $\beta_3/\beta_1$ and $v_d/v_F$, $\gamma_w$ and $\xi$ coincide,

\begin{equation}
 \gamma_w \approx \xi \approx \dfrac{m^2}{\hbar^4} \left[ 4v_F^2 + 8(v_d^2-v_F^2)\dfrac{\beta_3}{\beta_1} \right] \tau \beta_1^2.
\end{equation}
Asymptotically for $t\rightarrow \infty$, Eq.~\eqref{eq:PSHminusBroad} becomes $m_z^-(\bm{r}, t) \approx \rho(\bm{r},t) e^{-(\gamma_w-\xi)t}(1-\frac{\tau\mean{\Omega_y^2}}{2\xi})$. The factor $(1-\frac{\tau\mean{\Omega_y^2}}{2\xi}) < 0$ implies that the magnetization should flip at some instant, as we discuss below. The total relaxation rate in this asymptotic limit becomes

\begin{equation}
 \gamma_w - \xi \approx \dfrac{2m^2}{\hbar^4}(v_F^2 + 18v_d^2)\tau \beta_3^2 + \dfrac{1}{2\tau} \dfrac{(v_d \beta_3)^2}{(v_F \beta_1)^2},
\end{equation}
which is of the same order as the relaxation rates $\gamma_n$ and $\gamma_p$ of the narrow \PSHm and the \PSHp regimes, respectively.

The magnetization [see Fig.~\ref{fig:PSHminus-vsbroad}(f)] will have a single nodal line at a time $t = t_c$ set by the transcendental equation

\begin{align}
 \tanh(\xi t_c) &= \dfrac{2\xi}{\tau\mean{\Omega_y^2}}
             \approx 1 - \dfrac{\hbar^4}{8m^2 \tau^2} \dfrac{(v_d\beta_3)^2}{(v_F\beta_1)^4},
             \label{eq:nodetime}
\end{align}
where the approximate value is taken up to leading order in $v_d/v_F$ and $\beta_3/\beta_1$. For $v_d = 0$ there is no nodal line, i.e., $t_c \rightarrow \infty$.

For our set of parameters in Table \ref{tab:single} and $v_d = 30$~nm/ps we find $t_c \approx 115$~ps. However, the analytical solution above is only valid in the limit $\Gamma_0 \rightarrow \infty$. More precisely, this limit requires $|\tilde{\Xi}_x| \gg |\tilde{\Xi}_y|$, which yields

\begin{equation}
 \Gamma_0 \gg \Gamma_W = \dfrac{2\tau \mean{v_x\Omega_y}}{\mean{\Omega_x}} \approx 2\tau \dfrac{v_F^2\beta_1}{v_d\beta_3},
\end{equation}
with $\Gamma_W \approx 30$~$\mu$m. This is much wider than the typical laser spot used in recent experiments, where $\Gamma_0$ is $\sim$(sub)micron. Nonetheless, the numerical solutions for $\Gamma_0$ near the transition from the narrow to the wide \PSHm regimes are shown in Fig.~\ref{fig:PSHminus-vsbroad}. The single nodal line is already visible for $\Gamma_0 = \Gamma_C \gtrsim 2$~$\mu$m, while its instant $t_c$ strongly depends on $\Gamma_0$. For $\Gamma_0 > 30$~$\mu$m the numerical data matches $t_c = 115$~ps (not shown).

\subsection{Beyond the PSH regimes and general discussion}
\label{sec:Beyond}

To guide our discussion, let us consider a one-subband system similar to the sample discussed in Ref.~\onlinecite{altmann2016current}. The relevant parameters are shown in Table~\ref{tab:single}, for which we get the Fermi velocity $v_F \approx 274$~nm/ps, and the diffusion constant $D \approx 38$~$\mu\text{m}^2$/ps.

Starting with a $\delta$ packet, $m_z(\bm{r},0) = \delta(\bm{r})$, the numerical drift and diffusion pattern of $m_z(y,0)$ at $x=0$ is shown in Fig.~\ref{fig:narrowvsalpha} for $v_d = 30$~nm/ps and different values of $\alpha$. The exact \PSHp occurs in panel (e), while the exact \PSHm is shown in panel (a). These match Eq.~\eqref{eq:PSHplus} and Eq.~\eqref{eq:PSHminusNarrow}, respectively. 

For $\alpha \gtrsim -0.7$~meV\AA~ the stripes in the magnetization pattern are clearly visible. We can track the node lines to extract the wave number $\kappa_y$ and frequency $\omega$ to compare with the zeros of the cosine in Eq.~\eqref{eq:PSHplus}. These are shown in Fig.~\ref{fig:qw}. In the experiment of Ref.~\onlinecite{altmann2016current} the authors measure $\kappa_y$ and $\omega$ for a fixed $\alpha$ near the \PSHp regime and vary the electric field strength (or $v_d$). The numerical data in Fig.~\ref{fig:qw} show that the \PSHp solutions remain valid for a wide range of $\alpha$ around the exact \PSHp regime. Far from the \PSHp regime, near $\alpha = -0.7$~meV\AA~ both $\kappa_y$ and $\omega$ diverge as the stripes vanish. In Fig.~\ref{fig:qw} we calculate $\kappa_y$ and $\omega$ only for magnetization maps that have enough nodal lines to establish a periodicity ($\alpha > -0.7$~meV\AA), otherwise we set $\kappa_y = \omega = 0$ ($\alpha < -0.7$~meV\AA). 

The relaxation rate $\gamma$ is minimum ($1/\gamma$ is maximum) at the PSH regimes, as shown in Fig.~\ref{fig:gamma} for $v_d = 0$. There we compare $\gamma$ extracted from the numerical solutions of Fig.~\ref{fig:narrowvsalpha} with the analytical expressions of our PSH$^\pm$ approximate solutions, Eq.~\eqref{eq:gammap} and Eq.~\eqref{eq:gamman}. For any $v_d$, the strength of the \PSHp and the \PSHm peaks are similar. However, their position shifts away from the PSH$^\pm$ conditions, i.e., $\alpha = \pm (\beta_1-\beta_3)$, with increasing $v_d$. This new feature is more pronounced for the \PSHm regime, and cannot be neglected if one desires to explore this case experimentally.

The magnetization dynamics may strongly depend on the initial broadening $\Gamma_0$ of the packet, which is set by the laser spot of the pump beam. 
For the \PSHp regime, a wide packet solution can be extracted from the $\delta$ packet by convolution, as was done in Ref.~\onlinecite{altmann2016current}. 
In contrast, for initially wide packets, the dynamics of Eq.~\eqref{eq:PSHminusBroad} may dominate in the \PSHm regime. Figure \ref{fig:PSHminus-vsbroad} shows the transition between the narrow and wide \PSHm regimes from the numerical solutions of Eq.~\eqref{eq:diffmag}. 

In the narrow \PSHm regime ($\Gamma_0 \ll \Gamma_C$) the spin precession is static, given by $\cos(\kappa_x x)$ in Eq.~\eqref{eq:PSHminusNarrow}. For $x=0$, the magnetization is constant and one only observes the drift and diffusion process along $y$. For $\Gamma_C < \Gamma_0 < \Gamma_W$, the system is transitioning from the narrow to the wide regime. Within this range we only have numerical solutions, which qualitatively match the wide \PSHm regime ($\Gamma_0 > \Gamma_W$); i.e., the magnetization flips only once. As seen in Fig.~\ref{fig:PSHminus-vsbroad} for $\Gamma_0$ within the transition range, the nodal line moves to smaller $t$ with increasing $\Gamma_0$. It matches the wide \PSHm regime for $\Gamma_0 > \Gamma_W \sim 30$~$\mu$m (not shown).

The magnetization flip of the \PSHm regime requires a finite drift velocity; see Eq.~\eqref{eq:nodetime}. Here we always consider $v_d \ll v_F$, introducing the drift as a small shift of the Fermi circle. For the wide \PSHm regime, the drift velocity appears in $\mean{\Omega_x} \propto v_d$ (see Appendix \ref{app:pshm}), and affects $\xi = \frac{1}{2}\sqrt{\tau^2\mean{\Omega_y^2}-4\mean{\Omega_x}^2}$. For large $v_d$, the square root would become negative and $\xi$ purely imaginary. This indicates that for large $v_d$, the wide \PSHm regime would show oscillations and stripes as in the \PSHp regime. However, a large $v_d$ is not consistent with the RW model. Nonetheless, we interpret the single nodal line of the wide \PSHm regime as the first node of these speculative drift-induced oscillations.

\subsection{External Magnetic Field}
\label{sec:magnetic}

Consider the Zeeman term from an in-plane magnetic field $\bm{B} = (B_x, B_y, 0)$. It adds to the Hamiltonian as $H_Z = \frac{1}{2}g\mu_B \bm{B}\cdot\bm{\sigma}$, and complements the spin precession adding $\bm{\Omega}_B = g\mu_B\bm{B}/\hbar$ to $\bm{\Omega}$ in Eq.~\eqref{eq:OmegaSOC}, \textit{i.e.} $\bm{\Omega} \rightarrow \bm{\Omega} + \bm{\Omega}_B$. Here $g$ is the effective $g$ factor and $\mu_B$ is the Bohr magneton. Consequently, the magnetic field yields corrections to the $\Omega$-averages in Eq.~\eqref{eq:diffmag}; see Appendix \ref{app:magnetic}. Particularly, a finite $B_y$ makes $\mean{\Omega_x\Omega_y}$ and $\mean{v_y\Omega_y}$ finite, while the previous analytical PSH solutions require them to be zero. Therefore we shall focus on cases where only the $B_x$ component is finite, i.e. an in-plane magnetic field transverse to the drift velocity. 

For $\bm{B} = B_x \hat{x}$, the corrected $\Omega$ averages are 
$\mean{\Omega_x} \rightarrow \mean{\Omega_x} + \mean{\Delta \Omega_x}$,
$\mean{\Omega_x^2} \rightarrow \mean{\Omega_x^2} + \mean{\Delta \Omega_x^2}$,
$\mean{v_y\Omega_x} \rightarrow \mean{v_y\Omega_x} + \mean{\Delta v_y\Omega_x}$,
where the corrections are

\begin{align}
 \mean{\Delta\Omega_x} &= \dfrac{g\mu_B}{\hbar} B_x,\\
 \mean{\Delta\Omega_x^2} &= \left(\dfrac{g\mu_B}{\hbar} B_x\right)^2 + \dfrac{4m}{\hbar^2} \dfrac{g\mu_B}{\hbar} B_x v_d (\alpha+\beta_1-2\beta_3),\\
 \mean{\Delta v_y\Omega_x} &= v_d \dfrac{g\mu_B}{\hbar} B_x.
\end{align}

\subsubsection{\texorpdfstring{\PSHp and $B_x$}{PSH+ and Bx}}

For $gB_x > 0$, the $\Omega$-average corrections above will not affect the approximations used to get the analytical \PSHp solution. But for $gB_x < 0$, the corrections will reduce the intensity of $\tilde{\Xi}_x$, which may invalidate the condition $|\tilde{\Xi}_x| \gg |\tilde{\Xi}_y|$. In general, our \PSHp solution will hold for positive $gB_x$, and for small negative $gB_x$ that does not break the inequality. The corrected wave number $\kappa_y \rightarrow \kappa_y+\kappa_{B}$ and frequency $\omega \rightarrow \omega + \omega_B$ are

\begin{align}
 \kappa_B &= \dfrac{g\mu_B B_x}{\hbar}\dfrac{2v_d}{v_F^2 + 2v_d^2} \approx \dfrac{g\mu_B B_x}{\hbar} \dfrac{2v_d}{v_F^2} ,\\
 \omega_B &= \dfrac{g\mu_B B_x}{\hbar}\dfrac{v_F^2}{v_F^2 + 2v_d^2} \approx \dfrac{g\mu_B B_x}{\hbar}\left(1-\dfrac{2v_d^2}{v_F^2}\right).
\end{align}

For $v_d \ll v_F$, only $\omega$ is affected by $B_x$: the Zeeman frequency simply adds to the frequency of the cubic SOC. This can be understood because in the \PSHp situation the effective SOC magnetic field is mostly aligned with $B_x$. Therefore, $B_x$ leads to an additional tilt of the oscillation stripes, which has been used in Ref.~\onlinecite{walser2012direct} to determine the SOC.

\subsubsection{\texorpdfstring{\PSHm and $B_x$}{PSH- and Bx}}

In general, for a system near the \PSHm regime, a strong $B_x$ will invalidate the approximations used to obtain analytical solutions. However, a small $B_x$ can be used to manipulate the nodal line in the single magnetization flip of the wide \PSHm regime. The magnetic field adds a term to $\xi \rightarrow \xi + \xi_B$, with

\begin{multline}
 \xi_B \approx -\dfrac{\hbar^2}{8m^2}\dfrac{\beta_1+2\beta_3}{\tau v_F^2 \beta_1^3}(g\mu_B B_x)^2 \\
   + \dfrac{\hbar}{2m}\dfrac{v_d\beta_3}{\tau v_F^2\beta_1^2}g\mu_BB_x.
\end{multline}

For $v_d = 0$, the magnetic correction of $\xi$ modifies the transcendental equation for $t_c$ [Eq.~\eqref{eq:nodetime}], yielding

\begin{equation}
 \tanh(\xi t_c) \approx 1 - \dfrac{\hbar^6}{32 m^4\tau^2}\dfrac{ (g\mu_BB_x)^2}{v_F^4 \beta_1^4}.
\end{equation}
Therefore a small $B_x$ can play the role of the drift velocity and induce a magnetization flip for the wide \PSHm regime.

\section{Two subbands}
\label{sec:twosubbands}

The inter- and intrasubband SOCs were extensively studied in Refs.~\onlinecite{bernardes2006spin, EsmerindoPRL2007, calsaverini2008intersubband, dettwiler2017Stretchable, fu2015spin, Fu2015Skyrmion}, including a proposal for a crossed persistent spin helix \cite{Fu2015Skyrmion} (cPSH) and an intrinsic mechanism for edge spin accumulation \cite{khaetskii2016giant, Felix2013SpinHall}. In this section we investigate this cPSH within the RW model. The cPSH occurs when the subbands are set to opposite PSH regimes; e.g., the first subband is on the \PSHp, while the second is on the \PSHm regime. The magnetization profile of this crossed regime is not yet explored experimentally. 

Here we find two possible scenarios for the two-subband RW model. In the first case, Sec.~\ref{sec:weak2}, we consider the intersubband scattering (ISS) to be weak, such that the dynamics of the electrons of the first and second subband are independent. The resulting magnetization is an incoherent sum of the magnetization of each subband, and leads to a checkerboard pattern similar to the cPSH of Ref.~\onlinecite{Fu2015Skyrmion}. The second scenario, Sec.~\ref{sec:strong2}, corresponds to a regime of strong ISS. The random scattering events allow the electrons to quickly visit the Fermi circles of both subbands, allowing us to consider the subbands as semiclassical random variables. In this case each electron feels an average field that now includes an average over the subbands.

Particularly, we will discuss situations where one subband is near the \PSHp regime, while the other is near the \PSHm. This can occur in wide quantum wells, where the Hartree repulsion creates effective triangular wells with opposite slopes at each side of the heterostructure \cite{Felix2013SpinHall}. In Ref.~\onlinecite{Felix2013SpinHall} the symmetric and antisymmetric wave functions are nearly degenerate, allowing a rotation towards wave functions located on the left and right triangular wells. Another possibility is to have a slightly asymmetric well, breaking the degeneracy between left and right states.

The random walk model for two subbands will, in general, give finite values for all averages in Eq.~\eqref{eq:diffmag}. Consequently, the approximations presented for the single-subband cases will break. Moreover, the introduction of subband-dependent SOC, as well as intersubband SOC, leads to a large number of variables to analyze. Instead, for simplicity, the following discussion uses the representative parameters of Table \ref{tab:twosubbands}. 

\begin{table}[hb!]
\caption{Parameters for the two-subband system.}
\label{tab:twosubbands}
\begin{center}
\begin{ruledtabular}
\begin{tabular}{cll}
Parameter & Value & Description\\
\hline
$m$       & $0.067m_0$ & Effective mass (GaAs)\\
$n_s$     & $8\times 10^{11}$~cm$^{-2}$ & 2DEG density\\
$\Delta_{12} = 2\varepsilon_-$ & 7~meV & Subband energy splitting\\
$(n_1, n_2)$ & $(5.0, 3.0)\times 10^{11}$~cm$^{-2}$ & Density per subband\\
$\beta_{1,1} \approx \beta_{1,2}$ & 3.7~meV\AA & Linear Dresselhaus SOC\\
$(\beta_{3,1}, \beta_{3,2})$ & $(0.86, 0.52)$~meV\AA & Cubic Dresselhaus SOC\\
\multicolumn{2}{c}{$-5 \leq (\alpha_2=-\alpha_1) \leq 5$~meV\AA} & Rashba SOC\\
$\eta$ & $\pm 1$~meV\AA & intersubband SOC\\
$\Gamma$ & $\pm 1$~meV\AA & intersubband SOC\\
$\tau$    & 1~ps & Average scattering time\\
\end{tabular}
\end{ruledtabular}
\end{center}
\end{table}

\subsection{Subband and spin precession vectors}

The effective Hamiltonian \cite{Fu2015Skyrmion} for a two-subband 2DEG with SOC is $H = H_0 + H_{SOC}$, with

\begin{align}
 H_0 &= \left(\dfrac{\hbar^2 k^2}{2m} + \varepsilon_+\right) - \varepsilon_- \lambda_z,\\
 H_{SOC} &= \dfrac{\hbar}{2}\bm{\sigma}\cdot \left[  \bm{\Omega}_+ - \lambda_z \bm{\Omega}_- + \lambda_x \bm{\Omega}_{12}\right],
 \label{eq:Hsoc}
\end{align}
where $\varepsilon_{\pm} = (\varepsilon_2\pm\varepsilon_1)/2$ is defined in terms of the first ($\nu = 1$) and second ($\nu=2$) subband energies $\varepsilon_\nu$, $\bm{\sigma} = (\sigma_x, \sigma_y, \sigma_z)$ are the spin operators, similarly $\bm{\lambda} = (\lambda_x, \lambda_y, \lambda_z)$ act on the subband subspace, $\bm{\Omega}_\pm = (\bm{\Omega}_2\pm\bm{\Omega}_1)/2$, and $m$ is the effective mass. 

The spin-orbit fields for each subband $\nu$ and the intersubband field are

\begin{eqnarray}
 \bm{\Omega}_\nu &=& \dfrac{2}{\hbar}
 \begin{pmatrix}
  (+\alpha_\nu+\beta_{1,\nu})k_y + 2\beta_{3,\nu}\dfrac{k_x^2-k_y^2}{k^2}k_y\\
  (-\alpha_\nu+\beta_{1,\nu})k_x - 2\beta_{3,\nu}\dfrac{k_x^2-k_y^2}{k^2}k_x\\
  0
 \end{pmatrix}
 ,\\
 \bm{\Omega}_{12} &=& \dfrac{2}{\hbar}
 \begin{pmatrix}
  +(\eta-\Gamma) k_y\\
  -(\eta+\Gamma) k_x\\
  0
 \end{pmatrix}.
\end{eqnarray}
Here we consider the Rashba $\alpha_\nu$, linear $\beta_{1,\nu}$ and cubic $\beta_{3,\nu}$ Dresselhaus contributions for each subband $\nu = \{1, 2\}$, and the intersubband SOCs, $\eta$ and $\Gamma$. In general we shall consider $\bm{\Omega}_\nu$ and $\bm{\Omega}_{12}$ as perturbations, such that the energy dispersion remains approximately parabolic near the Fermi level. The $\bm{\Omega}$ vector that defines the spin precession frequency for the RW model in Eq.~\eqref{eq:diffmag} is now $\bm{\Omega} = \bm{\Omega}_+ - \lambda_z\bm{\Omega}_- + \lambda_x \bm{\Omega}_{12}$, which is coupled to the subband operators $\bm{\lambda}$.

If the subband energy difference $\Delta_{12} = \varepsilon_2-\varepsilon_1$ is large compared to the $H_{SOC}$ correction, we can use Löwdin perturbation theory to decouple the subbands. This is shown in Appendix \ref{app:lowdin}. Consequently, at each subband the electron spin feels an effective precession vector $\bm{\Omega}_{\nu}^\text{eff} = \bm{\Omega}_\nu + \bm{\Omega}_\nu^{(3)}$, where the small corrections $\bm{\Omega}_\nu^{(3)}$ are given by

\begin{multline}
\label{eq:mainO3x}
 \Omega^{(3)}_{\nu,x} = \dfrac{-\hbar^2}{4\Delta_{12}^2} \Bigg[\Omega_{{\bar{\nu}},x}[(\Omega_{12,x})^2-(\Omega_{12,y})^2] +
 \\
 +2\;\Omega_{{\bar{\nu}},y}\;\Omega_{12,x}\;\Omega_{12,y}\Bigg],
\end{multline}
\begin{multline}
\label{eq:mainO3y}
 \Omega^{(3)}_{\nu,y} = \dfrac{-\hbar^2}{4\Delta_{12}^2} \Bigg[\Omega_{{\bar{\nu}},y}[(\Omega_{12,y})^2-(\Omega_{12,x})^2] +
 \\
 +2\;\Omega_{{\bar{\nu}},x}\;\Omega_{12,x}\;\Omega_{12,y}\Bigg],
\end{multline}
where $\bar{\nu}$ refers to the complementary subband.

\subsection{Weak intersubband scattering}
\label{sec:weak2}

The ISS might be weak for large subband splitting $2\varepsilon_-$ and low temperatures, such that ISS events that require large momentum transfer are suppressed.
In this regime, sets of electrons initialized at different subbands constitute independent ensembles with well defined $\mean{\bm{\lambda}} = (0, 0, \pm 1)$. Each ensemble follows the dynamics of a single-subband, as in Sec.~\ref{sec:Single}. The total magnetization is then an incoherent sum of the magnetization $\bm{m}_\nu(\bm{r},t)$ from each occupied subband,

\begin{equation}
 \bm{m}(\bm{r},t) = \sum_\nu \bm{m}_\nu(\bm{r},t).
\end{equation}

\subsubsection{Crossed PSHs}

The dynamics of each $\bm{m}_\nu(\bm{r},t)$ depends on the parameters of subband $\nu$. A particularly interesting case is when one subband is at the \PSHp regime and the other is on the \PSHm regime. This leads to the crossed-PSH (cPSH) regime, or persistent skyrmion lattice (PSL), first discussed in Ref.~\onlinecite{Fu2015Skyrmion}. 

Consider the parameters from Table \ref{tab:twosubbands}, where subband $\nu=1$ is near the \PSHp regime with $\alpha_1 \approx \beta_{1,1} - \beta_{3,1}$, while the other subband $\bar{\nu}=2$ is near the \PSHm regime with $\alpha_{2} \approx -\beta_{1,2} + \beta_{3,2}$. 
The z-components of the magnetizations for each subband are 

\begin{align}
\label{eq:mag2plus}
  m_1(\bm{r},t) &= \rho(\bm{r},t) e^{-\gamma_{y,1} t} \cos(\kappa_{y,1} y + \omega_1 t),
 \\
\label{eq:mag2minus}
  m_{2}(\bm{r},t) &= \rho(\bm{r},t) e^{-\gamma_{x,2} t} \cos(\kappa_{x,2} x).
\end{align}

\begin{figure}[ht!]
 \centering
 \includegraphics[width=\columnwidth,keepaspectratio=true]{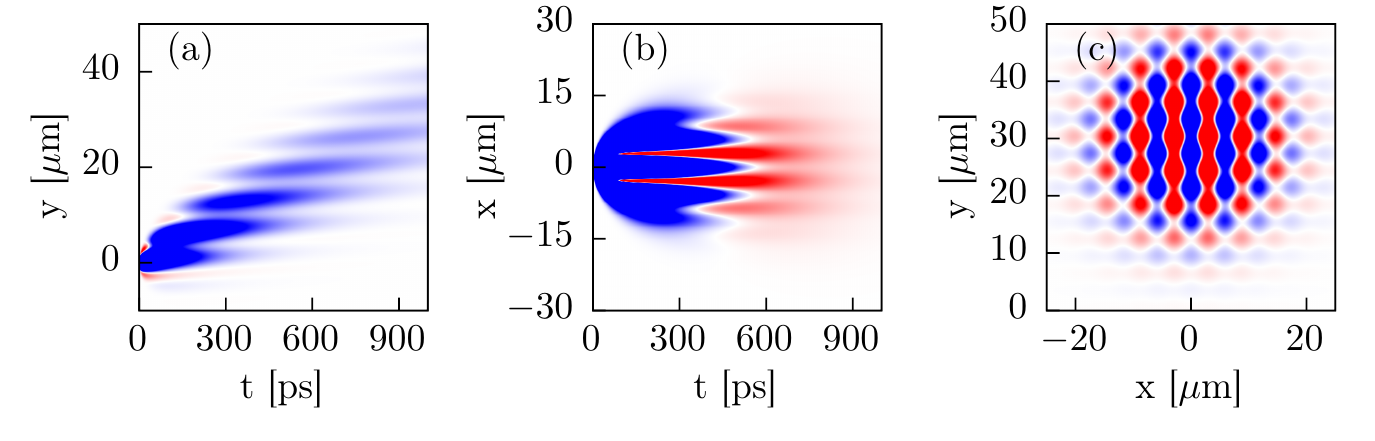}
 \caption{Magnetization patterns for two-subband crossed PSHs in the weak-ISS regime. The first (second) subband is in the \PSHp (\PSHm) regime. (a) At $x=0$ the magnetization profile is always positive, but shows the \PSHp stripes between zero and finite magnetization. (b) At $y=0$ the $x-t$ map shows a zigzag pattern. (c) Checkerboard pattern \cite{Fu2015Skyrmion} at $t = 1000$~ps centered at $y = v_d t$. In panel (b) the colors are highly saturated for better visualization of the pattern.}
 \label{fig:cPSHWeak}
\end{figure}

The resulting magnetization will have oscillations along $x$ and $y$, yielding the checkerboard pattern of the cPSH; see Fig.~\ref{fig:cPSHWeak}.
For $x=0$ the pattern on the $yt$ map is approximately given by $[1+\cos(\kappa_{y,1} y + \omega_1 t)]$, which renders the \PSHp stripes shifted to positive magnetization. This is a consequence of the incoherent superposition of the \PSHp of subband $\nu=1$, Eq.~\eqref{eq:mag2plus}, and the \PSHm of subband $\bar{\nu}=2$, Eq.~\eqref{eq:mag2minus}, which gives a positive background to the \PSHp stripes.
On the $xy$ map for fixed $t$, the superposition of oscillations along $x$ and $y$ leads to the checkerboard pattern in Fig.~\ref{fig:cPSHWeak}(c).

The drift velocity $v_d$ sets a finite $\omega_1$, which drives a motion of the checkerboard pattern with velocity $v_y = -\omega_1/\kappa_{y,1} \propto v_d$. Additionally, the drift velocity affects the relaxation rates of the \PSHp and \PSHm subbands differently, hence $\gamma_{y,1} \neq \gamma_{x,2}$ and for large $t$ one mode will prevail. In Fig.~\ref{fig:cPSHWeak}(c) this is seen as slight preference to form stripes rather than the checkerboard pattern at the center of the package.

\subsubsection{Intersubband SOC corrections}

The intersubband SOCs are introduced via the effective precession vector $\bm{\Omega}_\nu^{\text{eff}}$ (see Appendix \ref{app:lowdin}).
For overall weak SOC, the new terms in $\bm{\Omega}^\text{eff}_\nu$
do not break the approximations used to obtain the PSH$^\pm$ regimes. Consequently, the intersubband SOC simply introduces corrections to the wave vectors
$\kappa_{x,\bar{\nu}} \rightarrow \kappa_{x,\bar{\nu}} + \delta \kappa_{x,\bar{\nu}}$, $\kappa_{y,\nu} \rightarrow \kappa_{y,\nu} + \delta \kappa_{y,\nu}$,
and frequency 
$\omega_\nu \rightarrow \omega_\nu + \delta \omega_\nu$.
Up to leading order in $v_d/v_F$ they read

\begin{multline}
  \delta \kappa_{y,\nu} = -\dfrac{2m^3}{\Delta_{12}^2 \hbar^4} v_{f,\nu}^2
  \Big[
  \eta(-2\Gamma+\eta)\alpha_{\bar{\nu}} - \Gamma(\Gamma-2\eta)\beta_{1,\bar{\nu}} \\ + (\Gamma-\eta)^2\beta_{3,\bar{\nu}}
  \Big],
\end{multline}
\begin{multline}
  \delta \omega_\nu = -\dfrac{2m^3 v_F^2 v_d}{\Delta_{12}^2 \hbar^2}
  \Big[
  \eta  (\eta -2 \Gamma ) \alpha_{\bar{\nu}} + \Gamma  (\Gamma -2 \eta ) \beta_{1,\bar{\nu}} \\ -2 (\Gamma -\eta )^2 \beta_{3,\bar{\nu}}
  \Big],
\end{multline}

\begin{multline}
  \delta \kappa_{x,\bar{\nu}} = \dfrac{2m^3}{\Delta_{12}^2 \hbar^4} v_{f,\bar{\nu}}^2
  \Big[
  \eta(2\Gamma+\eta)\alpha_{\nu} - \Gamma(\Gamma+2\eta)\beta_{1,\nu} \\ + (\Gamma+\eta)^2\beta_{3,\nu}
  \Big].
\end{multline}
Additionally, the intersubband SOC will lead to corrections to the relaxation rates ($\gamma_n$, $\gamma_p$ and $\gamma_w$) and frequency $\xi$. However, these are large expressions that we choose not to show explicitly.

\subsection{Strong intersubband scattering}
\label{sec:strong2}

For the strong-ISS regime we consider that both intra- and intersubband scattering times are comparable, and both are much shorter than the spin precession period. Therefore, the random walk process allows the electron to wander throughout the Fermi circles of all occupied subbands. Consequently, here we include an average over the subbands into the $\mean{\cdots}$ averages of the RW model. For a generic term $\mathcal{O}_\nu(\theta)$, the average now reads

\begin{equation}
 \mean{\mathcal{O}} = \dfrac{1}{2\pi N_\nu}\sum_{\nu=1}^{N_\nu} \int_0^{2\pi} \mathcal{O}_\nu(\theta) = \dfrac{1}{N_\nu} \sum_{\nu=1}^{N_\nu} \mean{\mathcal{O}}_\nu,
 \label{eq:averages2}
\end{equation}
where $N_\nu$ is the number of occupied subbands. The short form on the right-hand side expresses the subband average contracting the $\theta$ average into $\mean{\mathcal{O}}_\nu$.

The $\theta$ averages $\mean{\mathcal{O}}_\nu$ are equivalent to the ones of the single-subband cases, but calculated with $\bm{\Omega}_\nu^\text{eff}$, which introduces subband-dependent parameters ($\alpha_\nu$, $\beta_{1,\nu}$, $\beta_{3,\nu}$ and $k_{F,\nu}$), as well as the intersubband couplings ($\eta$ and $\Gamma$). For a drift velocity along $y$, these averages remain null $\mean{\Omega_y} = \mean{\Omega_x\Omega_y} = \mean{v_x\Omega_x} = \mean{v_y\Omega_y} = 0$. The others can be easily calculated algebraically, but now yield long expressions (not shown); namely, these are $\mean{\Omega_x}$, $\mean{v_x \Omega_y}$, $\mean{v_y \Omega_x}$, $\mean{\Omega_j^2}$, for $j=\{x,y\}$. Overall, the precession is dominated by $\mean{\bm{\Omega}} = \frac{1}{2}(\mean{\bm{\Omega}}_1 + \mean{\bm{\Omega}}_2)$, yielding subband-averaged SOCs $\alpha_+ = \frac{1}{2}(\alpha_1+\alpha_2)$, $\beta_{1,+} = \frac{1}{2}(\beta_{1,1}+\beta_{1,2})$, and $\beta_{3,+} = \frac{1}{2}(\beta_{3,1}+\beta_{3,2})$, plus perturbative corrections.

Considering the parameters of Table \ref{tab:twosubbands}, we find that the RW averages are nearly isotropic for $v_d = 0$ and $\alpha_1 = -\alpha_2 = 0$, with $\mean{\bm{\Omega}} = 0$, $\mean{v_x \Omega_y} \lesssim \mean{v_y \Omega_x}$, and $\mean{\Omega_x^2} \lesssim \mean{\Omega_y^2}$.
The strict isotropic dynamics would be equivalent to the pure Dresselhaus case discussed for a single-subband in Ref.~\onlinecite{Stanescu2007PhysRevB}, where the magnetization follows a Bessel pattern given by

\begin{equation}
 m_z(r,t) \propto \dfrac{e^{-\gamma_0 t}}{\sqrt{t}}J_0(\kappa_0 r),
\end{equation}
where the wave number $\kappa_0$ and the relaxation rate $\gamma_0$ are

\begin{align}
  \kappa_0 &= \dfrac{2\mean{v_y \Omega_x}}{\mean{v_y^2}},\\
  \gamma_0 &= \dfrac{3\tau}{2}\mean{\Omega_x^2} - \dfrac{\tau\mean{v_y\Omega_x}^2}{\mean{v_y^2}}.
\end{align}

\begin{figure}[ht!]
 \centering
 \includegraphics[width=\columnwidth,keepaspectratio=true]{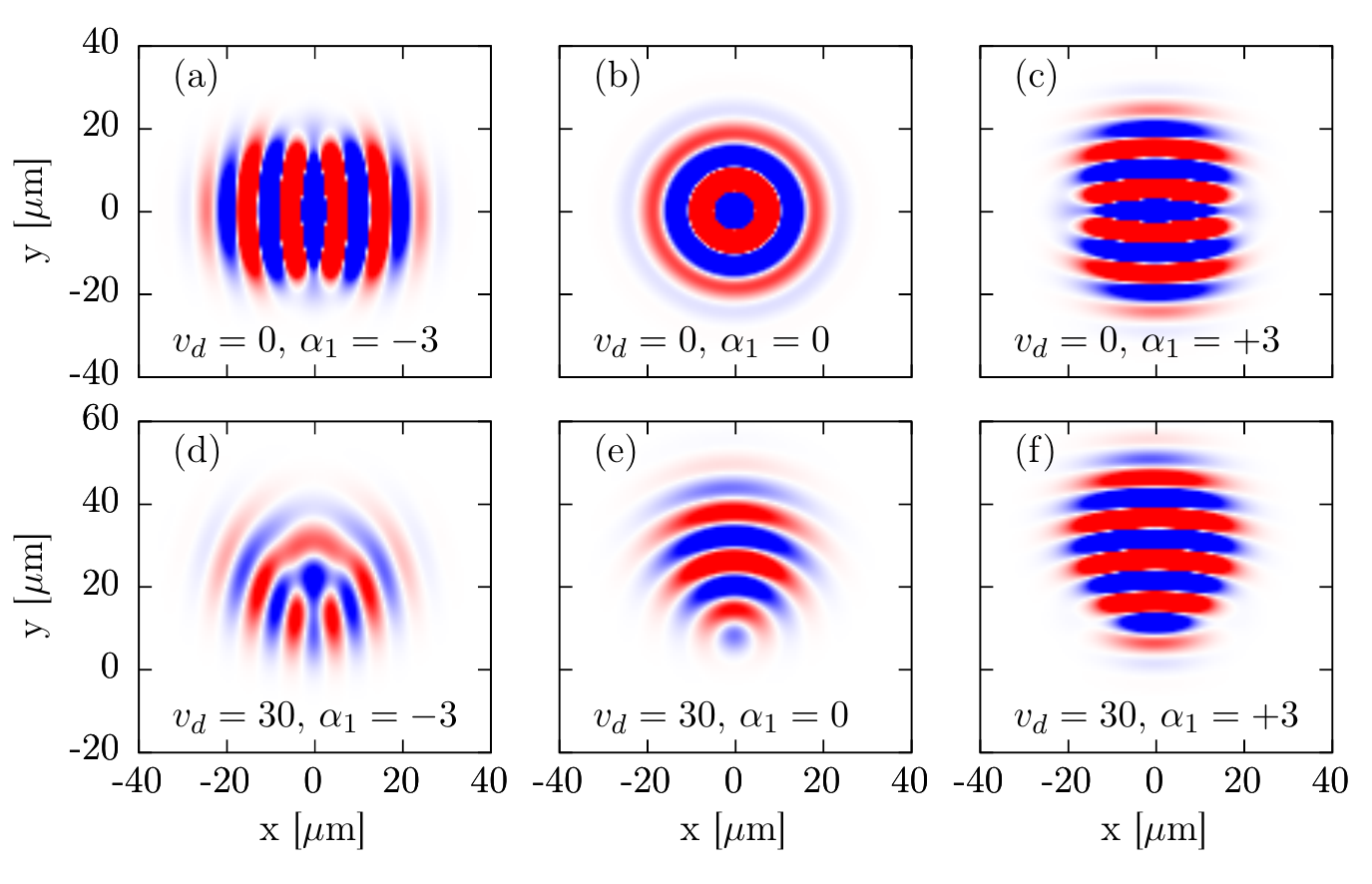}
 \caption{
 Magnetization patterns for the two-subband crossed PSH in the strong-coupling limit.
 The drift velocity $v_d$ and the subband-dependent Rashba SOC $\alpha_1 = -\alpha_2$ are indicated in each panel in units of nm/ps and meV\AA, respectively. The maps are taken at $t=1000$~ps and other parameters are set in Table \ref{tab:twosubbands}. The isotropic case in panel (b) matches the Bessel solutions. For finite $\alpha_1$ the Bessel pattern is distorted. A finite $v_d$ induces the drift of the packet (envelope), but the inner pattern moves with a slow velocity.
 Due to the large relaxation rate $\gamma_0$ of the Bessel solutions, the colors here had to be strongly saturated for clear visualization of the patterns.
 }
 \label{fig:cPSHStrong}
\end{figure}

For $\alpha_1 =-\alpha_2= 0$ and $v_d =0$ the Bessel pattern is shown in Fig.~\ref{fig:cPSHStrong}(b) and matches well the approximate isotropic solution above. Beyond the isotropic limit (i.e., for finite but small $\alpha_1 = -\alpha_2$ and $v_d$), the precession pattern still resembles the isotropic Bessel solutions. A finite $v_d$ drives the Gaussian envelope drift along $y$, but the inner magnetization pattern drifts with a slower velocity, as seen in Figs.~\ref{fig:cPSHStrong}(d)-(f). An equivalent effect was discussed previously for the single-subband \PSHp case, where the precession pattern $\propto \cos(\kappa_y y + \omega t)$ yields a pattern velocity $-\omega/\kappa_y$. A finite $\alpha_1 = -\alpha_2$ distorts the pattern vertically [Figs.~\ref{fig:cPSHStrong}(a) and \ref{fig:cPSHStrong}(d)] or horizontally [Figs.~\ref{fig:cPSHStrong}(c) and \ref{fig:cPSHStrong}(f)], for $\alpha_1 > 0$ and $\alpha_1 < 0$, respectively. 

Since the dynamics is dominated by the subband-averaged SOCs, even though the subbands are set to the cPSH regime, the averaged SOCs deviate from this regime. Indeed $\alpha_+ = 0$ for $\alpha_1 =-\alpha_2$, while $\beta_{1,+} \neq 0$. Consequently, the strong-ISS regime for the cPSH setup shows a short lifetime $\gamma_0^{-1} \sim 0.1$~ns, much shorter than the single-subband PSH regimes ($\sim 1$~ns in Fig.~\ref{fig:gamma}).

\section{Final Remarks and Conclusions}
\label{sec:conclusion}

\subsection{Limitations of the Random Walk model}

The RW model provides an elegant description of the spin diffusion process. However, there are limitations. To go from the symbolic definition of the joint probability, Eq.~\eqref{eq:joint}, to the differential equation for the magnetization, Eq.~\eqref{eq:diffmag}, we have performed a Taylor expansion for small $\Delta\bm{r} = \bm{v}_n\tau$ and $\Delta\bm{s}_n$. Additionally, we use $\bm{s}_{n+1}$ from Eq.~\eqref{eq:ssol} in Eq.~\eqref{eq:mint}. We combine these keeping only terms that are linear in $\bm{s}$; otherwise it is not possible to recover the definition of the magnetization, Eq.~\eqref{eq:mint}, and write the differential equation for $\bm{m}(\bm{r},t)$. As shown here, this approximation is remarkably good for samples with small SOC coefficients, like GaAs quantum wells \cite{altmann2016current}. However, for strong SOC one cannot neglect the spin-orbit locking that couples the spin with the (Fermi) velocity. This condition would lead to a spin-dependent $\Delta\bm{r}$ (\textit{zitterbewegung} \cite{Schliemann2005Zitter, Schliemann2005ZitterB, bernardes2006spin, Schliemann2007SideJump}), thus breaking the conditions required to recover the magnetization in the expansion approach. With a different approach, in Ref.~\onlinecite{Sinova2011StrongSOC} the authors consider the strong-SOC regime, and show that the PSH arises from Rabi oscillations. 
Additionally, the RW model does not account for spin-charge coupling, which is discussed in Refs.~\onlinecite{Stanescu2007PhysRevB,Kleinert2007DriftB1} for single-subband systems.

\subsection{Conclusions}

We have analyzed the spin diffusion dynamics in two-subband systems, extending the random walk model to account for the subband dynamics. Our model includes the Rashba, linear and cubic Dresselhaus, and the intersubband spin-orbit couplings. Additionally, we have discussed the effects of initial packet broadening and external magnetic fields.

For the dynamics of two-subband systems, two possible scenarios were identified regarding the ISS rates.
For weak ISS, the subbands are effectively uncoupled and the magnetization dynamics is essentially an average of the magnetization of the individual subbands (incoherent sum). Consequently, if the subbands are set into the crossed-PSH regime, both magnetizations will show a long lifetime, resulting in the checkerboard pattern \cite{Fu2015Skyrmion}. We show that for a finite drift velocity, the single-subband relaxation rates for the \PSHp and \PSHm are different. Therefore, in the weak-ISS regime, a finite drift velocity could lead to different relaxation rates, such that for large $t$ only one of the magnetizations will prevail, returning to the single-subband striped pattern.
For strong ISS, we have seen that the fast subband dynamics introduces subband-averaged spin-orbit couplings, rather than a subband-averaged magnetization (as in the weak-ISS case). Consequently, even if the individual subbands are set into the crossed-PSH regime, their averaged SOC will not be close to a PSH. Instead, we obtain a nearly isotropic Bessel pattern with short lifetime.

Spintronic applications require long spin lifetimes. From the results presented, this can be achieved in two-subband systems by setting individual subbands into PSH regimes if the ISS is sufficiently weak. However, for strong ISS, the subband-averaged SOC is the main character. It may destroy the long-lived cPSH. However, one would still recover a long lifetime if the subband-averaged SOCs fall close to the PSH regime. Therefore, the extension to two-subband systems provides an additional handle to fine-tune the dynamics to obtain long lifetimes.

\begin{acknowledgments}
We acknowledge financial support from the Brazilian agencies CNPq, CAPES, and FAPEMIG. G.J.F. thanks Jiyong Fu for helpful discussions. F.G.G.H. acknowledges financial support from Grant No. 2014/25981-7 of the São Paulo Research Foundation (FAPESP). P.A. and G.S. acknowledge financial support from the NCCR QSIT of the Swiss National Science Foundation.
\end{acknowledgments}

\appendix

\section{Magnetic field along z}
\label{app:omegaz}

In the main text we have assumed $\Omega_z = 0$ to express Eq.~\eqref{eq:diffmag} in a simple form. However, a finite $\Omega_z$ could be introduced by an external magnetic field $\bm{B} = B_z\hat{z}$, in which case one must add $W_z$ to the matrices in Eq.~\eqref{eq:diffmag}. Namely

\begin{equation}
 W_z =
 \begin{pmatrix}
  -\tau \mean{\Omega_z^2} & -\Xi_z & +\tau\mean{\Omega_x\Omega_z}\\
  \Xi_z & -\tau\mean{\Omega_z^2} & \tau\mean{\Omega_y\Omega_z}\\
  \tau\mean{\Omega_x\Omega_z} & \tau\mean{\Omega_y\Omega_z} & 0
 \end{pmatrix},
\end{equation}
where $\Xi_z = \mean{\Omega_z} - 2\tau\Big[ \mean{v_x\Omega_z}\partial_x + \mean{v_y\Omega_z}\partial_y \Big]$, and $\Omega_z = \frac{g\mu_B}{\hbar}B_z$.

\section{Expressions for the averages and other secondary quantities mentioned in the text}

Here we show large or cumbersome expressions that are not relevant for the main discussion.

\subsection{Averages for the single-subband case without magnetic field}
\label{app:single}

We assume that the drift velocity is along $\hat{y}$. The null averages were already mentioned in the main text, 
$\mean{\Omega_y} = \mean{\Omega_x \Omega_y} = \mean{v_x\Omega_x} = \mean{v_y\Omega_y} =  0$.
The finite ones are

\begin{equation}
 \mean{v_x^2} = \dfrac{1}{2}v_F^2,
\end{equation}

\begin{equation}
 \mean{v_y^2} = \dfrac{1}{2}v_F^2 + v_d^2,
\end{equation}

\begin{equation}
 \mean{\Omega_x} = \dfrac{2m}{\hbar^2} v_d \left(\alpha+\beta_{1} - 2\dfrac{v_d^2+v_F^2}{v_F^2}\beta_{3} \right),
\end{equation}

\begin{align}
 \nonumber
 \mean{v_y \Omega_x} = \dfrac{m}{\hbar^2}\Bigg[& (2v_d^2 + v_F^2)(\alpha + \beta_{1}) \\
                                               &- \left(v_F^2 + 10v_d^2 + 4 \dfrac{v_d^4}{v_F^2}\right)\beta_{3} \Bigg],
\end{align}

\begin{equation}
 \mean{v_x \Omega_y} = -\dfrac{m}{\hbar^2}v_F^2\Bigg(\alpha-\beta_1+\beta_3-2\dfrac{v_d^2}{v_F^2}\beta_3\Bigg).
\end{equation}

The expressions for $\mean{\Omega_x^2}$ and $\mean{\Omega_y^2}$ are large; therefore here we choose to show only their series expansion up to second order in $v_d/v_F$,

\begin{multline}
 \mean{\Omega_x^2} \approx \dfrac{2m^2 v_F^2}{\hbar^4}
 \Bigg[
       \Big( \left(\alpha +\beta_1\right){}^2-2 \left(\alpha +\beta_1\right)\beta_3+2 \beta_3^2 \Big)\\
       + 2\dfrac{v_d^2}{v_F^2}\Big( \left(\alpha +\beta_1\right){}^2-10\left(\alpha +\beta_1\right)\beta_3 +18 \beta_3^2 \Big)
 \Bigg]
\end{multline}

\begin{multline}
  \mean{\Omega_y^2} \approx \dfrac{2m^2 v_F^2}{\hbar^4}
  \Bigg[
	\left(\alpha -\beta_1\right){}^2+2\left(\alpha -\beta_1\right)\beta_3+2 \beta_3^2 \\
	- \dfrac{4v_d^2}{v_F^2} \left(\alpha-\beta_1 \right)\beta_3
  \Bigg]
\end{multline}

\subsection{\texorpdfstring{\PSHp: $\gamma_p$, $\kappa_y$ and $\omega$}{PSH+}}
\label{app:pshp}

\begin{equation}
 \gamma_p = \tau\Bigg(\mean{\Omega_x^2} + \dfrac{1}{2}\mean{\Omega_y^2} - \dfrac{\mean{v_y\Omega_x}^2}{\mean{v_y^2}}\Bigg),
\end{equation}

\begin{equation}
 \kappa_y = \dfrac{\mean{v_y\Omega_x}}{\mean{v_y^2}},
\end{equation}

\begin{equation}
 \omega = \mean{\Omega_x} - \kappa_y v_d
\end{equation}

\subsection{\texorpdfstring{\PSHm: $\gamma_n$, and $\kappa_x$}{PSH-}}
\label{app:pshm}

For the narrow initial packet, $\Gamma_0 \ll \Gamma_C$:

\begin{equation}
 \gamma_n = \tau\left(\mean{\Omega_y^2}+\dfrac{\mean{\Omega_x^2}}{2}-\dfrac{\mean{v_x\Omega_y}^2}{\mean{v_x^2}}\right) + \dfrac{1}{2\tau}\dfrac{\mean{\Omega_x}^2}{\mean{\Omega_y^2}-\mean{\Omega_x^2}},
\end{equation}

\begin{equation}
 \kappa_x =  \dfrac{\mean{v_x\Omega_y}}{\mean{v_x^2}}
\end{equation}

For the wide initial packet, $\Gamma_0 \gg \Gamma_C$:

\begin{equation}
 \gamma_w = \tau \left( \mean{\Omega_x^2} + \dfrac{1}{2}\mean{\Omega_y^2} \right)
\end{equation}

\begin{equation}
 \xi = \dfrac{1}{2}\sqrt{\tau^2\mean{\Omega_y^2}^2 - 4\mean{\Omega_x}^2}
\end{equation}

\subsection{Complement to the single-subband averages due to an external magnetic field}
\label{app:magnetic}

A finite in-plane magnetic field $\bm{B} = (B_x, B_y, 0)$ introduces additional terms to the $\Omega$-averages of Sec.~\ref{app:single}. Labeling the additive terms of each $\Omega$-average with a $\Delta$, e.g., $\mean{\bm{\Omega}} \rightarrow \mean{\bm{\Omega}} + \mean{\Delta \bm{\Omega}}$, the extra terms due to $\bm{B}$ up to leading order in $v_d/v_F$ are

\begin{equation}
 \mean{\Delta\bm{\Omega}} = \dfrac{g\mu_B}{\hbar}\bm{B},
\end{equation}

\begin{equation}
 \mean{\Delta\Omega^2_x} = \left(\dfrac{g\mu_B}{\hbar}B_x\right)^2 + \dfrac{4m}{\hbar^2}\dfrac{g\mu_B}{\hbar}B_x v_d(\alpha+\beta_1-2\beta_3),
\end{equation}

\begin{equation}
 \mean{\Delta\Omega^2_y} = \left(\dfrac{g\mu_B}{\hbar}B_y\right)^2,
\end{equation}

\begin{multline}
 \mean{\Delta\Omega_x \Omega_y} = \left(\dfrac{g\mu_B}{\hbar}\right)^2 B_xB_y \\ + \dfrac{2m}{\hbar^2} \dfrac{g\mu_B}{\hbar}B_y v_d(\alpha+\beta_1-2\beta_3),
\end{multline}

\begin{equation}
 \mean{\Delta v_x\bm{\Omega}} = 0,
\end{equation}

\begin{equation}
 \mean{\Delta v_y\bm{\Omega}} = v_d \dfrac{g\mu_B}{\hbar} \bm{B}.
\end{equation}

\section{Effective model for each subband}
\label{app:lowdin}

To decouple the subbands we use the Löwdin perturbation theory. Consider $H = H_0 + H'$, with the perturbation given by the SO term $H' = H_{SOC}$. Up to second order in $(\varepsilon_\nu-\varepsilon_{\bar{\nu}})^{-1}$ we obtain an effective Hamiltonian $\tilde{H}_\nu$ for each subband $\nu =  \{1, 2\}$,

\begin{equation}
 \tilde{H}_\nu = \left(\dfrac{\hbar^2 k^2}{2m} + \varepsilon_\nu\right) + \Delta^{(2)}_{\nu}
 + \dfrac{\hbar}{2}\bm{\sigma}\cdot(\bm{\Omega}_\nu + \bm{\Omega}^{(3)}_\nu),
\label{eq:Heff}
\end{equation}
where $\Delta^{(2)}_{\nu}$ are the spin-independent corrections to the subband energy, 
and $\bm{\Omega}^{(3)}_\nu = \Omega^{(3)}_{\nu,x}\hat{x}+\Omega^{(3)}_{\nu,y}\hat{y}$ are the corrections for the effective magnetic field. Using $\nu$ and ${\bar{\nu}}$ to refer to opposite subbands, and taking the approximation $E \approx \varepsilon_\nu + \hbar^2k^2/2m$, we get

\begin{equation}
 \Delta^{(2)}_{\nu} = \dfrac{\Big(\hbar \bm{\Omega}_{12}\Big)^2}{4(\varepsilon_\nu-\varepsilon_{\bar{\nu}})},
\end{equation}

\begin{multline}
\label{eq:O3x}
 \Omega^{(3)}_{\nu,x} = \dfrac{-\hbar^2}{4\Delta_{12}^2} \Bigg[\Omega_{{\bar{\nu}},x}[(\Omega_{12,x})^2-(\Omega_{12,y})^2] +
 \\
 +2\;\Omega_{{\bar{\nu}},y}\;\Omega_{12,x}\;\Omega_{12,y}\Bigg],
\end{multline}

\begin{multline}
\label{eq:O3y}
 \Omega^{(3)}_{\nu,y} = \dfrac{-\hbar^2}{4\Delta_{12}^2} \Bigg[\Omega_{{\bar{\nu}},y}[(\Omega_{12,y})^2-(\Omega_{12,x})^2] +
 \\
 +2\;\Omega_{{\bar{\nu}},x}\;\Omega_{12,x}\;\Omega_{12,y}\Bigg],
\end{multline}
where $\Delta_{12} = \varepsilon_2-\varepsilon_1$ is the energy difference between the subbands.

This effective model can be used to account for the neglected inter-subband SOC effects in Secs.~\ref{sec:weak2} and \ref{sec:strong2}.
For each subband $\nu$, the effective precession vector from Eq.~\eqref{eq:Heff} is

\begin{equation}
 \bm{\Omega}_\nu^\text{eff} = \bm{\Omega}_\nu + \bm{\Omega}^{(3)}_\nu.
 \label{eq:effOmega}
\end{equation}
Additionally, the subband energy dispersion remains approximately spin-independent, $E_\nu = \varepsilon_\nu + \Delta_\nu^{(2)} + \hbar^2 k^2/2m$. This assures that the Fermi velocity is isotropic and spin-independent, as required by the RW model of Sec.~\ref{sec:rwmodel}.

\bibliography{refs}

\end{document}